\documentclass[sigconf]{acmart}

\usepackage{listings}
\usepackage{xcolor}
\usepackage{xspace}
\usepackage{graphicx}
\usepackage{enumitem}
\usepackage{algorithm}
\usepackage{algpseudocode}
\usepackage{subfig}
\usepackage{tikz}
\usepackage{multirow}
\usepackage{seqsplit}
\usepackage{siunitx}

\newcommand{\codename}{\mbox{\textsc{SimGuard}}\xspace}
\newcommand{\hash}[1]{{\ttfamily\seqsplit{#1}}}

\definecolor{lightgray}{rgb}{0.95,0.95,0.95}

\newcommand*\circled[1]{\tikz[baseline=(char.base)]{
            \node[shape=circle,fill,inner sep=1pt] (char) {\textcolor{white}{#1}};}}

\lstdefinelanguage{Solidity}{
  keywords={ contract, function, if, else},
  sensitive=true,
  comment=[l]{//},
  morecomment=[s]{/*}{*/},
  morestring=[b]"
}

\AtBeginDocument{%
  }

\setcopyright{acmlicensed}
\copyrightyear{2018}
\acmYear{2018}
\acmDOI{XXXXXXX.XXXXXXX}
\acmConference[Conference acronym 'XX]{Make sure to enter the correct
  conference title from your rights confirmation email}{June 03--05,
  2018}{Woodstock, NY}
\acmISBN{978-1-4503-XXXX-X/2018/06}

\begin{document}

\title{Blockchain Transaction Simulation Phishing}


\author{Xiaocan Wang}
\affiliation{%
  \institution{Stevens Institute of Technology}
}
\email{xwang221@stevens.edu}

\author{Shixuan Guan}
\affiliation{%
  \institution{Stevens Institute of Technology}
}
\email{sguan6@stevens.edu}

\author{Tong Yang}
\affiliation{%
  \institution{Rutgers University}
}
\email{ty339@scarletmail.rutgers.edu}

\author{Xiapu Luo}
\affiliation{%
  \institution{The Hong Kong Polytechnic University}
}
\email{csxluo@comp.polyu.edu.hk}

\author{Yue Duan}
\affiliation{%
  \institution{Singapore Management University}
}
\email{yueduan@smu.edu.sg}

\author{Kai Li}
\affiliation{%
  \institution{Stevens Institute of Technology}
}
\email{kli50@stevens.edu}

\renewcommand{\shortauthors}{Trovato et al.}

%




\begin{abstract}
Cryptocurrency users have increasingly become targets of phishing and scam attacks. To mitigate these threats, leading crypto wallets (e.g., MetaMask) have introduced \emph{transaction simulation}, which previews a transaction’s balance changes before on-chain execution. While effective against traditional fund-draining attacks, we show that this defense can itself be exploited by a new phishing technique, which we term \emph{transaction simulation phishing}. This attack uses carefully-crafted smart contracts whose execution depends on dynamic blockchain state, causing simulations to display benign or profitable outcomes while the actual on-chain execution redirects users’ funds to attacker-controlled addresses.

We present the first comprehensive study of transaction simulation phishing. We first develop a taxonomy of phishing contracts that can be utilized to facilitate this attack. Then, we propose \codename, a bytecode-level detection system that combines static and dynamic program analysis to identify phishing contracts. Applying \codename to Ethereum, Binance Smart Chain, Avalanche, and Polygon, we detect over 4{,}000 phishing contracts deployed between August 2024 and June 2025. Our analysis identifies more than 5{,}700 victims and approximately \$3.48 million USD in losses, 91.5\% of which occurred on Ethereum. Moreover, our clustering result reveals that the largest phishing contract cluster alone accounts for about 83\% of the total losses. These results expose a critical weakness in current wallet defenses and highlight the urgent need for more robust transaction simulation mechanisms.
\end{abstract}

\maketitle
\pagestyle{plain}

\section{Introduction}
The rapid advancement of blockchain technology has stimulated substantial growth in the global cryptocurrency market. Unfortunately, this growth has been accompanied by a surge in cybercrime, including phishing and scam attacks that exploit multiple layers of the blockchain ecosystem. Prior studies have shown that attackers abuse smart contracts~\cite{me:knownattacks,perez2021smart,gao2020tracking,kell2021forsage,xia21scams}, crypto wallets~\cite{ChengHLZLLR19,he2023txphishscope,yan2024stealing,guan24ccs,tsuchiya2025blockchainaddresspoisoning,chen2025dissecting,he2025phishing,meisami2025sigscope}, and social media platforms~\cite{xia20covidscams,vakilinia2022cryptocurrency,xigao2023doublenothing,li2023understanding,li2023towards} to deceive victims and steal cryptocurrency assets. The financial impact is severe, with losses in the first half of 2025 alone exceeding \$2.17 billion USD~\cite{loss2025}. These trends underscore the urgent need for effective countermeasures to protect cryptocurrency users.

Crypto wallets serve as the primary interface through which users manage and transfer digital assets and therefore play a critical role in safeguarding user funds. To mitigate phishing attacks, wallet developers have widely adopted \emph{transaction simulation}, a feature that has grown in popularity in recent years. Leading wallets such as MetaMask~\cite{me:metamask}, Coinbase~\cite{me:coinbase}, and Trust Wallet~\cite{me:trust} now integrate transaction simulation modules that speculatively execute a transaction before it is submitted on-chain. By previewing the expected balance changes, transaction simulation enables users to assess whether a transaction’s outcome aligns with their expectations, thereby improving transparency and security. Consequently, many existing phishing attacks~\cite{he2023txphishscope,li2023towards,yan2024stealing,meisami2025sigscope,chen2025dissecting} that directly drain users’ funds can be effectively exposed during simulation, significantly reducing their success rates.

\textbf{New phishing threat:} Despite its defensive intent, transaction simulation has inadvertently introduced a new attack surface. In this paper, we demonstrate that attackers have exploited transaction simulation itself to deceive wallet users, a novel phishing tactic that we term \emph{transaction simulation phishing}. This attack leverages carefully-crafted phishing contracts whose execution semantics depend on dynamic properties of the blockchain environment. During simulation, such contracts can exhibit benign and profitable outcomes, inducing users to approve the transaction. However, after submission, changes in blockchain state or transaction context cause the on-chain execution path to diverge from the simulated one, which redirects the user’s funds to attacker-controlled addresses and results in financial losses to them.

\textbf{Taxonomy and detection system:} Motivated by this emerging phishing threat, we conduct a comprehensive study of transaction simulation phishing and develop techniques to detect phishing contracts at scale. Based on the insight that any dynamic element of the blockchain execution environment may be exploited to manipulate a contract's execution flow, we systematically analyze blockchain variables that influence transaction semantics and construct a taxonomy of six classes of transaction simulation phishing contracts: \emph{storage-control}, \emph{external-control}, \emph{Gas-control}, \emph{Gasprice-control}, \emph{blocknumber-control}, and \emph{timestamp-control}. Building on this taxonomy, we design \codename, a bytecode-level detection system capable of detecting these phishing contracts at large-scale across EVM-compatible blockchains. At a high level, \codename first symbolically executes contract bytecode to enumerate all possible execution traces and then identifies suspicious fund transfer behaviors based on the characteristic of transaction simulation phishing, where one branch returns funds and apparent profits to the caller while another redirects funds to external addresses under the attacker's control. To minimize false positives, \codename further validates suspicious contracts via runtime testing in a controlled execution environment.

\textbf{Detection results and research findings:} We apply \codename to four major EVM-compatible blockchains, including Ethereum mainnet, Binance Smart Chain (BSC), Avalanche (Avax), and Polygon, and identify a total of 4,224 phishing contracts, including 480 on Ethereum, 293 on BSC, 3,316 on Avax, and 315 on Polygon. Our analysis shows that \emph{Gas-control} contracts dominate in prevalence, while Avax hosts the largest number of deployed phishing contracts. The timeline analysis reveals that transaction simulation phishing first emerged in August 2024 and intensified between March and May 2025, with over 100 contracts deployed in a single day across the four blockchains. By tracing the on-chain transaction histories of each detected phishing contract, we identify more than 5,700 victim addresses suffering losses exceeding \$3.48 million USD, with 91.5\% occurring on Ethereum. Among all phishing types, our analysis indicates that \emph{storage-control} contracts are the most profitable, accounting for over \$2.97 million USD (85.34\%) of total losses, which is primarily due to their characteristics in offering a deterministic control over the execution outcome. Our further analysis reveals that although the number of daily victims remained high between March and May 2025, the most severe losses were concentrated in the early stages of the campaign, including a single day in January 2025 with losses exceeding \$1 million USD. Moreover, our investigation shows that attackers also achieve extraordinarily high returns, with return-on-investment (ROI) ranging from 4,463\% to 83,495\% across four blockchains. Finally, our clustering result shows that a small number of coordinated attacker groups dominate the phishing campaign, with the largest cluster on Ethereum alone accounting for approximately \$2.88 million USD (83\% of total profits). Collectively, these findings demonstrate the severity, scalability, and economic incentives of transaction simulation phishing and highlight the urgent need for more robust wallet defenses.

Overall, our work makes the following contributions:
\begin{itemize}[leftmargin=*]
\item \textbf{Comprehensive analysis:} We present the first comprehensive study of \emph{transaction simulation phishing}, an emerging threat that specifically exploits transaction simulation mechanisms in modern crypto wallets.
\item \textbf{Taxonomy of phishing contracts:} We systematically analyze dynamic blockchain properties and construct a taxonomy of six classes of phishing contracts exploiting transaction simulation.
\item \textbf{A new detection system:} We design and implement \codename, a bytecode-level detection system that integrates symbolic execution, phishing-specific transfer recognition, and runtime verification, achieving zero false positives in the evaluation results on a ground-truth dataset.
\item \textbf{Large-scale measurement:} Applying \codename to four major blockchains, we identify over 4,000 phishing contracts, more than 5,700 victims, and approximately \$3.48 million USD in losses. We further reveal temporal trends and coordinated attacker clusters. We have open-sourced \codename and released the collected dataset to support future research\footnote{\url{https://anonymous.4open.science/r/Blockchain-Transaction-Simulation-Phishing-BE81}}.
\end{itemize}

\section{Background}
\label{sec:background}
\subsection{Ethereum and Smart Contract} 
Ethereum is one of the most widely adopted Proof-of-Stake (PoS) blockchain platforms and a pioneer in supporting general-purpose smart contracts. The smart contract feature has enabled a wide range of applications, including Decentralized Finance (DeFi), NFTs, gaming, and supply chain management. On Ethereum, users deploy and execute smart contracts via transactions, which update the contract’s state on the blockchain. Contract execution is handled by the Ethereum Virtual Machine (EVM), which runs on every node in the peer-to-peer network to replicate execution and confirm results. The EVM is a stack-based virtual machine with a standardized set of opcodes~\cite{me:evm}, including arithmetic, storage, and control flow instructions that operate on a last-in, first-out (LIFO) stack. Each opcode manipulates stack values, and complex contract logic emerges from these sequential operations. While contracts execute as EVM bytecode, developers typically write them in higher-level languages such as Solidity~\cite{me:sol}, which are then compiled into bytecode for deployment.


\subsection{EVM-compatible Blockchains}
Inspired by the success of Ethereum, people have also adopted the EVM standard in other blockchains, including Binance Smart Chain (BSC), Avalanche (Avax), and Polygon. Unlike Ethereum, BSC~\cite{me:bsc} offers faster block intervals and reduced transaction costs through its Proof-of-Staked Authority (PoSA) mechanism. Avax~\cite{me:avax} adopts a multi-chain architecture and a custom consensus protocol designed for low latency and high throughput. Polygon~\cite{me:polygon} provides a scaling framework that offers high-speed and cost-efficient transactions. Despite the differences in the consensus protocols and transaction fee mechanisms, all of them are EVM-equivalent and retain full compatibility with the smart contract programming language, Solidity. In other words, the same smart contract can be seamlessly deployed on all these EVM-compatible blockchains to produce equivalent execution results. 

\begin{figure}[!htbp]
\hspace{-0.5cm}
     \centering
     \includegraphics[width=1.0\linewidth]{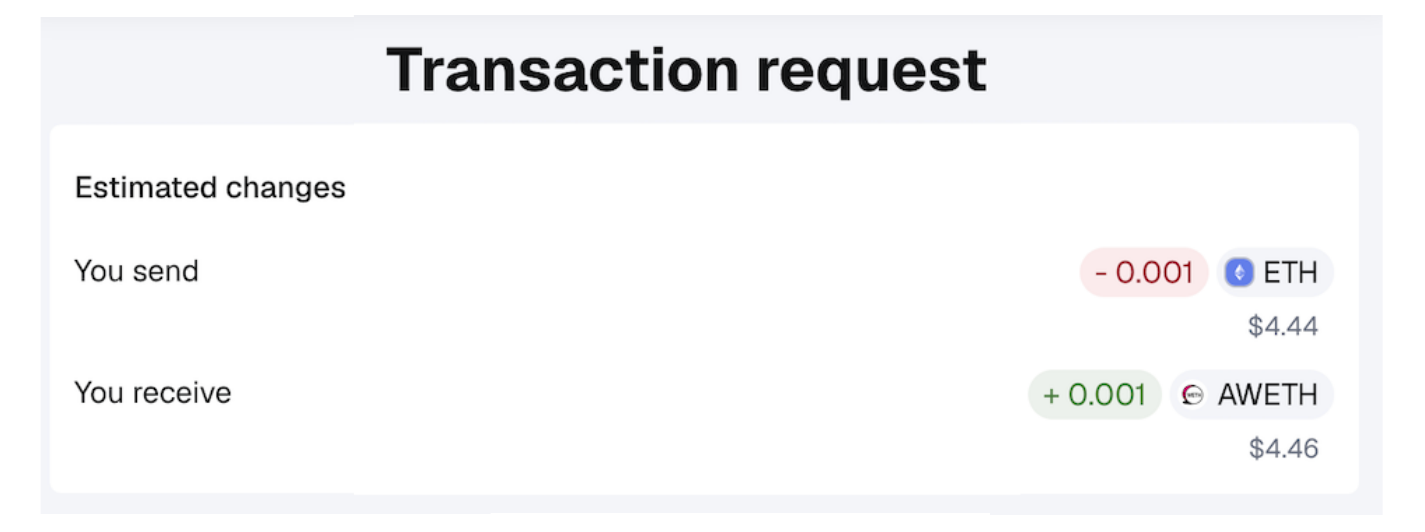}
     \caption{The transaction simulation feature on MetaMask.}
     \label{fig:sim}
\end{figure}

\begin{figure*}[!h]
    \centering
    \includegraphics[width=1.0\linewidth]{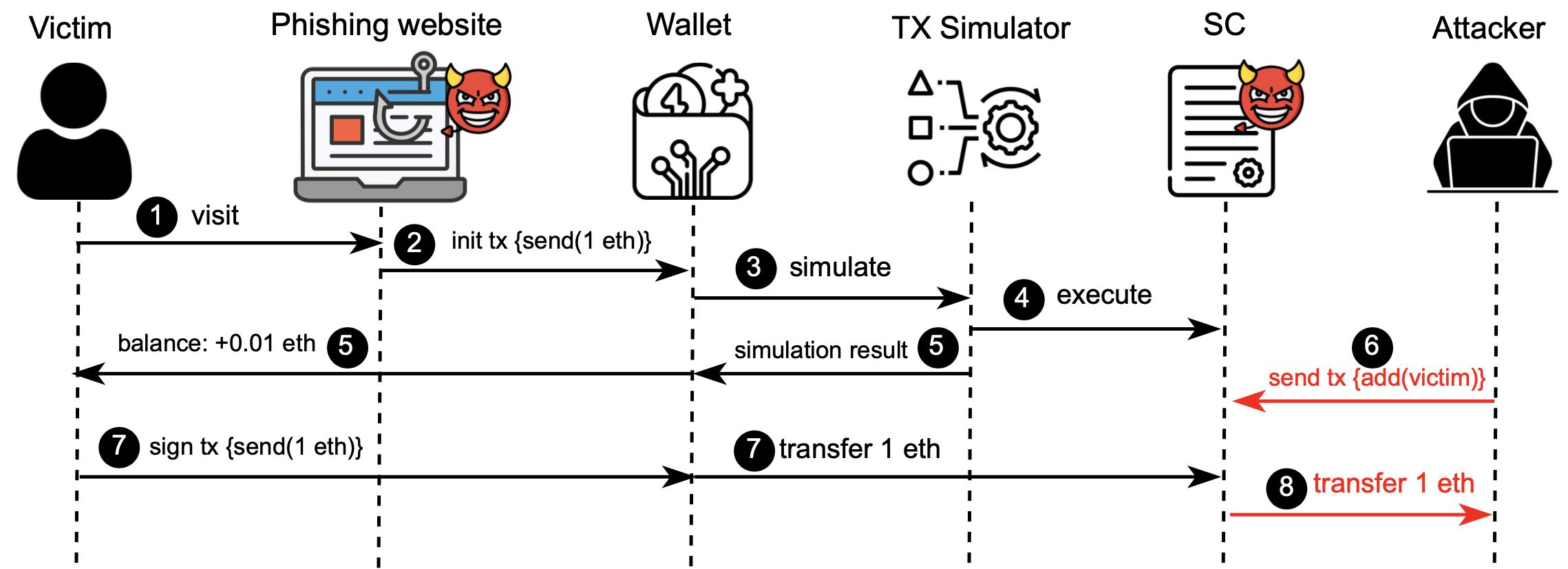}
    \caption{The workflow of the transaction simulation phishing: The victim first visits a cryptocurrency airdrop website that will connect to the victim's crypto wallet and initiate a transaction to deposit funds to a phishing contract. The wallet then forwards the transaction to the integrated transaction simulator for simulation. The simulation result will show that the victim's balance will be increased. The victim then signs and submits the transaction to the blockchain. Meanwhile, the attacker sends another transaction to frontrun the victim's transaction to manipulate the phishing contract's state, which alters the victim transaction's execution and redirects the victim's deposits to the attacker.}
    \label{fig:threat}
\end{figure*}
\subsection{Crypto Wallet and Transaction Simulation}
On blockchains, the user's cryptocurrency account is also referred to as external owned address (EOA), which is used to hold and manage cryptocurrency assets. To help users manage the cryptocurrency accounts, various crypto wallets are available in the App market. Due to the blockchain’s anonymity and irreversibility natures, crypto wallet users are increasingly targeted by various attacks~\cite{ChengHLZLLR19,yan2024stealing,meisami2025sigscope} and phishing scams~\cite{he2023txphishscope,yan2024stealing,guan24ccs,tsuchiya2025blockchainaddresspoisoning,chen2025dissecting,li2023towards}, leading to significant financial losses. To protect users against the threat, leading crypto wallets such as MetaMask, Coinbase, and Trust have integrated a transaction simulation module, which would executes the users' transaction based on the current blockchain state before they sign and submit it to the blockchain. As shown in Fig.~\ref{fig:sim}, when a transaction request is initiated on the MetaMask wallet, a preview of the transaction's simulation outcome is presented to users, which shows the balance change to its account. The goal of transaction simulation is to enhance the transaction's security and transparency so that users can verify whether the transaction execution result aligns with their expectations.


\section{Transaction Simulation Phishing}
\label{sec:phishing}
A key motivation for integrating transaction simulation is to reduce wallet users’ exposure to phishing and scam attacks. By previewing a transaction’s execution outcome, wallets can reveal unexpected balance reductions or unknown transfer destinations, prompting users to scrutinize and potentially reject suspicious transaction requests. While this mechanism is effective against traditional phishing attacks that directly drain users’ funds, we find that the ongoing arms race between attackers and defenders has led to a new attack vector. Specifically, transaction simulation itself introduces a new attack surface that can be abused by carefully-crafted phishing contracts to produce plausible profitable outcomes during simulation, thereby disguising malicious transactions as benign. We refer to this emerging attack vector as \emph{transaction simulation phishing}. In the following, we first present a concrete example demonstrating how this attack exploits transaction simulation, and then introduce a taxonomy of phishing contracts that enable such behavior.

\subsection{A Motivating Example: Storage-control Phishing Contract}
Fig.~\ref{fig:threat} illustrates the workflow of a transaction simulation phishing attack. As reported~\cite{defihacklabs}, the attack begins with a phishing website promoting a fraudulent "free cryptocurrency" campaign. The attackers lure users to visit the website via social media (e.g., X.com) (step~\protect\circled{1}). The website presents a "\emph{claim}" button that, when clicked, triggers a wallet connection request. Once the user approves the connection, the website initiates a transaction to transfer the user's funds (e.g., ETH) to a phishing smart contract (step~\protect\circled{2}). Before signing, the wallet forwards the transaction to its integrated transaction simulator (step~\protect\circled{3}), which executes it against the current blockchain state and generates a simulated execution result (step~\protect\circled{4}). The wallet then displays a preview showing the expected post-transaction balance change (step~\protect\circled{5}). To deceive the user, the phishing contract is crafted to return the deposited funds plus a tiny profit during simulation, making the balance appear increased. Meanwhile, the attackers submit a separate transaction (step~\protect\circled{6}) that manipulates the phishing contract’s state, such as blacklisting the user’s address. This alters the transaction’s actual execution path, redirecting the users' deposits to an attacker-controlled address. The wallet fails to detect this concurrent state change and does not update the simulation. Consequently, when the user approves and submits the transaction on-chain (step~\protect\circled{7}), the deposited funds are immediately transferred to the attacker (step~\protect\circled{8}), causing financial loss.

To better illustrate the attack, we show a representative phishing contract in Listing~\ref{lst:sample}. The contract has a \texttt{claim()} function that will be invoked by the victim user's transaction. This function contains an \texttt{if--else} branch whose execution depends on whether the caller's address appears in a blacklist. If the caller is not blacklisted, the branch at line 6 is executed, which returns the transferred funds along with a small additional reward intended to entice the victim. When the crypto wallet simulates the victim's transaction, the simulation result would indicate an apparent profitable outcome as the caller's address is not yet blacklisted. However, because the phishing website has already connected to the victim's wallet and obtained its address, the attacker can immediately submit a separate transaction to invoke \texttt{add(victim)} and add the victim's address to the blacklist. When the victim subsequently submits the original transaction to execute the \texttt{claim()} function on-chain, the contract's state has changed. As a result, the execution enters the alternative branch at line 8, which redirects the transferred funds to an attacker-controlled address. It can be seen that this attack exploits the classic time-of-check-to-time-of-use (TOCTOU) race condition between transaction simulation and on-chain execution. The time window between simulation and execution allows the attacker to stealthily manipulate the contract state, causing the actual execution semantics to diverge from the simulated result. Consequently, users who trust the wallet's simulation results are deceived into approving transactions that ultimately result in financial loss rather than the anticipated profit. 

\begin{lstlisting}[language=Solidity,caption={A sample storage-control phishing contract.}, captionpos=b, label={lst:sample}]
contract phishing {   
  function add(addr) public {
    blackList.push(addr);}
  function claim() public {
    if (msg.sender not in blackList){
     msg.sender.transfer(msg.value+0.0001 ether);}
    else{
     attacker.transfer(msg.value);}}
 }
\end{lstlisting}

The phishing contract described above uses \emph{contract storage} as the control variable to determine the transfer destination. By exploiting the dynamic and mutable nature of smart contract storage state, the attacker manipulates the contract's execution flow to deceive crypto wallet users. We therefore denote the contract as a \textbf{storage-control} phishing contract.

\begin{figure*}[!htbp]
    \centering
    \includegraphics[width=0.95\linewidth]{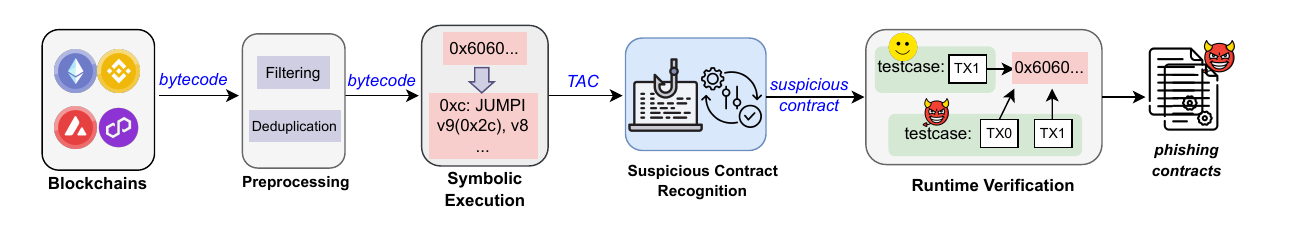}
    \caption{The phishing contract detection pipeline of \codename.}
    \label{fig:system}
\end{figure*}
\subsection{Taxonomy of Phishing Contracts}
\label{sec:taxonomy}
In principle, any dynamic property of the blockchain execution environment can be abused as a control variable in the phishing contracts. To ensure comprehensive coverage, we systematically analyze all blockchain property variables to identify those that can feasibly be exploited to control a contract's execution path. The property variables in blockchain fall into two broad categories~\cite{me:blockprop}: \emph{transaction variables} and \emph{block variables}. For transaction variables, the phishing website would typically set \texttt{from}, \texttt{to}, and \texttt{value}, \texttt{Gas}, \texttt{Gasprice}, and \texttt{data} fields when constructing the transaction. However, the transaction simulator (or user) could adjust the \texttt{Gas} and \texttt{Gasprice} fields~\cite{me:fordefi}. Such flexibility thus makes them possible control variables in the phishing contracts. While the \texttt{value} field could also be adjusted by users, since phishing attackers tend to drain the victim's entire balance, which varies across victims. Hence, it is unlikely to be selected as the control variable, as setting an appropriate threshold to \texttt{value} would be challenging. For block variables, we identify \texttt{block.number} and \texttt{block.timestamp} as practical control variables, as they are increased monotonically and therefore provide attackers with deterministic control over the execution path. In contrast, other block variables, such as \texttt{block.basefee} and \texttt{block.coinbase}, do not exhibit predictable behavior, making them less suitable for this phishing purpose. A detailed justification for each variable is provided in Appendix~\ref{sec:justify}. Based on this analysis, in addition to \textbf{storage-control} phishing contracts, we identify five additional variants of phishing contracts, as illustrated below.

\textbf{External-control:} This contract is similar to \textit{stotage-control}, except that the blacklist is maintained in an external contract. This contract references the external contract to determine whether the caller exists in the blacklist. If not, the contract returns the caller's deposits along with a tiny amount of profit. Otherwise, the contract redirects the deposits to the attacker. 
     
\textbf{Gas-control:} This contract uses the transaction's Gas limit (e.g., \texttt{tx.gas}) as the control variable. If below a predefined threshold (e.g., 5 million), the contract rewards the caller with a tiny amount of profit. Otherwise, the caller's deposits are moved to the attacker. The phishing website sets a large Gas limit (e.g., 6 million) in the initiated transaction. However, the transaction simulator may use a smaller default Gas limit to simulate the transaction, producing a profitable outcome. If the user approves the original transaction with the large Gas limit, its deposits are sent to the attacker. 


\textbf{Gasprice-control:} This contract is similar to \textit{Gas-control}, except that the Gas price (e.g., \texttt{tx.gasprice}) is used as the control variable. If it is below a threshold (e.g., 1000 Gwei), the contract transfers funds back to the caller. The phishing website sets a large Gas price (e.g., 1100 Gwei) in the transaction, while the transaction simulator may use a smaller default Gas price to simulate the transaction, producing a profitable outcome. Similarly, if the user approves the original transaction, the funds are sent to the attacker. 


\textbf{Blocknumber-control:} This contract implements a malicious \texttt{claim(uint256 number)} using \texttt{block.number} as the control variable. The contract compares \texttt{block.number} with the supplied \texttt{number} supplied in the transaction. If \texttt{block.number} is smaller, the contract transfers funds back to the caller. Otherwise, the funds are moved to the attacker. The phishing website sets \texttt{number} to the next block height (i.e., \texttt{block.number} + 1) to invoke \texttt{claim()} in the transaction. During simulation, the condition holds and the simulator produces a profitable outcome. However, when the transaction is executed on-chain, multiple blocks could have been produced, causing \texttt{block.number} to exceed \texttt{number}. As a result, the victim's deposits are sent to the attacker. 


\textbf{Timestamp-control:} This contract works similarly to the above except that it uses \texttt{block.timestamp} as the control variable. If \texttt{block.timestamp} is smaller than the supplied \texttt{timestamp} in the transaction, the contract transfers funds back to the caller. Otherwise, the deposits are sent to the attacker. The phishing website sets \texttt{timestamp} slightly ahead of the victim's local time (e.g., by adding a few seconds) to invoke \texttt{claim()}. During simulation, the condition evaluates to true. However, when the transaction is executed on-chain, multiple blocks could have been produced and \texttt{block.timestamp} exceeds the supplied \texttt{timestamp}, causing an altered execution flow that sends the victim's deposits to the attacker.


\section{Detection System: \codename}
\label{sec:system}
With the above taxonomy of phishing contracts, our work aims to detect them in real-world blockchains at a large scale. However, this is a challenging task, as most smart contracts do not provide source code, especially those containing malicious intent~\cite{abdelaziz2023smart}. In this work, we propose \codename to detect phishing contracts directly from the bytecode.

\subsection{Technical Design and Implementation}
The detection pipeline of \codename is presented in Fig.~\ref{fig:system}. At a high level, for smart contracts deployed on the EVM-compatible blockchains, \codename first leverages symbolic execution to convert the bytecode into the three-address-code (TAC) representation~\cite{me:tac}. Then, \codename parses the contract's TAC to detect suspicious transfer operations based on the characteristics of each phishing contract variant. Finally, contracts with suspicious transfer patterns will be further validated through dynamic analysis, which involves sending testing transactions to confirm whether their runtime behaviors align with the phishing activity. Overall, \codename is fully automated and can be applied to detect phishing contracts in EVM-compatible blockchains.

\textbf{Data collection and preprocessing:} Our detection pipeline begins with collecting smart contract bytecode from the historical transactions on the EVM-compatible blockchains. We leverage the Ethereum-ETL~\cite{me:etl} tool to retrieve the detailed transactions and the associated receipts, which are further parsed to extract the bytecode of deployed contracts and the deployed addresses. After collecting the deployed contracts and the bytecode, we then preprocess the collected dataset by filtering out benign contracts (e.g., popular ERC-20 tokens, DEXs) that are well-known. Since the blockchain may have duplicate contract code deployed, we further deduplicate the collected bytecode to reduce our analysis overhead.       


\textbf{Symbolic execution and suspicious contract recognition:} After preprocessing, \codename leverages the existing tool, greed~\cite{gritti2023confusum,ruaro2024crush}, to symbolically execute each remaining bytecode, producing a three-address-code (TAC) Intermediate Representation (IR). The symbolic execution tool takes the contract's bytecode as input and traverses all possible branches to generate various execution traces, which are saved into the TAC form consisting of simple instructions, each with at most three operands. After obtaining the TAC, \codename then detect suspicious transfer patterns based on the characteristics of each phishing contract variant. First of all, a common behavior of all phishing contracts is that they include multiple branches to execute the \texttt{transfer} operation. In one branch, funds (caller's deposits + additional amount) are returned to the caller. In other branches, the caller's deposits are sent to another address. Based on such a characteristic, we detect suspicious transfer patterns from the TAC using the following criteria:

\begin{itemize}[leftmargin=*]
\item In one basic code block, there is a \texttt{JUMPI} instruction connecting two successor basic blocks that both contain a \texttt{CALL} operation.
\item In one successor basic block, the \texttt{CALL} sends funds to \texttt{CALLER}, with the transferred amount greater than the caller's deposit (\texttt{CALLVALUE}).
\item In the other successor basic block, the \texttt{CALL} sends the caller's deposit (\texttt{CALLVALUE}) to an external address, which is loaded from the contract's storage via \texttt{SLOAD} or returned by another \texttt{CALL}.
\end{itemize}

\noindent With the above criteria, \codename will identify contracts with suspicious transfer patterns, with one returning the deposit as well as additional amounts to the caller, and the other moving the caller's deposit to another address. After identifying such suspicious contracts, \codename further analyzes them by tracing the data flow of the control variable referenced in the \texttt{JUMPI} instruction. If one of the below conditions meets, the suspicious contract is classified as the associated phishing contract variant. 
\begin{itemize}[leftmargin=*]
\item \textbf{Storage-control}: data flow dependent on both \texttt{SLOAD} and \texttt{CALLER}. 
\item \textbf{External-control}: data flow dependent on both \texttt{CALLDATA} and \texttt{CALLER}. 
\item \textbf{Gas-control}: data flow dependent on \texttt{tx.gas}.
\item \textbf{Gasprice-control}: data flow dependent on \texttt{tx.gasprice}.
\item \textbf{Blocknumber-control}: data flow dependent on \texttt{block.number}.
\item \textbf{Timestamp-control}: data flow dependent on \texttt{block.timestamp}.
\end{itemize}

\noindent Using the above detection criteria, \codename can identify highly suspicious contracts. \codename implements the detection criteria as an algorithm with the pseudo code provided in Appendix~\ref{sec:algo}.

\begin{table*}[!htbp]
\caption{Evaluation result of \codename on the ground-truth datasets.}
\label{tab:eval}
\resizebox{\textwidth}{!}{
\begin{tabular}{cccc|cccccccc}
\hline
\multicolumn{4}{c|}{\textbf{Ground-truth}}                                                                                            & \multicolumn{7}{c}{\textbf{\codename}}                                                & \multicolumn{1}{l}{}         \\ \hline
\multicolumn{2}{c|}{Phishing Contracts (P)}                                    & \multicolumn{2}{c|}{Benign Contracts (N)}            & \multicolumn{6}{c|}{Detection Result}                             & \multicolumn{2}{c}{Avg. Time (s)} \\ \hline
\multicolumn{1}{c|}{Snowball-sampling} & \multicolumn{1}{c|}{Gemini Synthesis} & \multicolumn{1}{c|}{Forta}   & Source-code Expansion & TP & FP & TN      & FN & Precision & \multicolumn{1}{c|}{Recall}  & Phishing          & Benign                       \\ \hline
\multicolumn{1}{c|}{14}                & \multicolumn{1}{c|}{30}               & \multicolumn{1}{c|}{139,451} & 23,524                & 33 & 0  & 162,975 & 1  & 100\%     & \multicolumn{1}{c|}{97.72\%} & 4.38              & 6.50                         \\ \hline
\end{tabular}
}
\end{table*}
\textbf{Runtime verification:} For each highly suspicious contract, \codename confirms whether its runtime execution aligns with the expected phishing behavior by sending transactions to execute it in a local environment using Foundry~\cite{me:foundry}. We send two groups of transactions, where the first group simulates the profit scenario and the second group simulates the financial loss scenario. For example, to test "storage-control" contracts, the first group only includes one transaction depositing funds to the contract. The expected behavior is that the contract will return the deposit as well as additional amounts, representing the profit scenario. The second group sends two transactions, with the first manipulating the storage state of the contract (e.g., invoking the \texttt{add()} function to blacklist the caller) and the second depositing funds to the contract. The expected behavior is that the contract will redirect the funds to another address, representing the financial loss scenario. Similarly, to test other phishing contract variants, we respectively test the profiting and financial loss scenarios by manipulating the storage state of the external contract or changing "\texttt{tx.gas}", "\texttt{tx.gasprice}", "\texttt{number}", and "\texttt{timestamp}" accordingly. To accurately trigger the profit and financial loss scenarios, we utilize the constraint solver~\cite{me:greedsolver} to construct the correct parameters in the testing transactions. For instance, to test "storage-control" phishing contracts, we locate \texttt{CALL} instructions that return funds to the caller (the profit scenario) and set that location as the target. Then, the constraint solver would resolve the required parameters for the transaction's execution to reach that location. To test the financial loss case, we then set the target location to statements manipulating the contract storage (e.g., \texttt{SSTORE}). For multiple solutions found, we will exhaust all of them in the testing transactions to ensure no phishing contracts will be missed. Finally, if both profiting and financial loss behaviors are observed, we confirm that the tested contract belongs to a phishing attacker.

\subsection{Evaluation of \codename}
This section evaluates \codename with a ground-truth dataset. 

\textbf{Phishing contracts:} Transaction simulation phishing is a newly emerging threat, so currently, no public ground-truth datasets exist. We construct our own dataset from a publicly reported phishing contract~\cite{defihacklabs} and use it as a seed contract to expand the dataset with a snowball-sampling approach. Specifically, we identify phishing contracts through three complementary strategies: (1) tracing contracts deployed by the same address and collecting those with similar bytecode; (2) querying Etherscan's Similar Contract API~\cite{me:similarAPI} to retrieve contracts similar to the reported one; and (3) identifying contracts that internally transfers to the same address and retaining those with similar bytecode. The bytecode similarity is measured by the existing method~\cite{me:slithersm} with a similarity threshold of 90\%. For each identified contract, we search its label on Etherscan and include only those explicitly labeled as "phishing" in the dataset. Using this snowball-sampling approach, we found 14 \emph{storage-control} phishing contracts. Because no labeled contracts were found for the remaining four phishing types, we synthesized additional phishing contracts to ensure coverage. Specifically, we instruct Gemini~\cite{me:gemini} to implement phishing contracts following the specifications described in Sec.~\ref{sec:taxonomy}. In addition to straightforward implementations (e.g., a single \texttt{if-else} branch), we also ask Gemini to incorporate various obfuscation techniques, including multiple branches, loops, dummy functions, and complex arithmetic expressions, to obscure the transfer logic. We thereby obtain five contracts for each phishing contract variant. In total, our phishing dataset consists of 44 contracts, spanning all six categories in our taxonomy.

\textbf{Benign contracts:} There is a public benign contract dataset released by Forta~\cite{me:fortaset}, which comprises 139,451 contracts that were obtained with a machine learning predictive model from the Ethereum mainet~\cite{me:fortaml}. Since the dataset was published three years ago and may not represent the current state, we expanded this dataset by looking into contracts deployed from Jan. 2024 to Aug. 2025 that have the source code published on Etherscan without a "phishing" label. In addition, we only consider contracts that have at least 100 transaction interactions. We then parse each contract's source code and classify them into the following three categories: (1) have no "transfer" operations; (2) have "transfer" operations but none of them are conditional; (3) have at least one conditional "transfer" operation. Through this process, we respectively expanded 8,145, 14,350, and 1,029 benign contracts in each category. In total, our benign dataset includes 139,451 + 23,524 = 162,975 contracts.        

\textbf{Evaluation results:} Table~\ref{tab:eval} summarizes the evaluation results of \codename. Among the 44 phishing contracts in our dataset, \codename successfully detects 43 of them. The only false negative is a synthesized \emph{Gasprice-control} phishing contract employing multiple obfuscation techniques, including opaque predicates, redundant loops, and indirect branch routing. These obfuscations prevent \codename from correctly tracing the data dependency between the \texttt{JUMPI} instruction and the \texttt{Gasprice} variable, leading to a missed detection. Nevertheless, \codename still shows the ability to detect the vast majority of phishing contracts (including all 14 real-world contracts labeled on Etherscan), achieving a recall of over 97\%. On the two benign contract datasets, \codename produces no false positives, resulting in 100\% precision. In addition, the evaluation result also suggests that \codename incurs a small execution overhead, with an average analysis time of 4.3 seconds on each phishing contract and 6.5 seconds on each benign contract.

\section{Phishing Contracts In the Wild}
\label{sec:result}
This section presents the detection results and an in-depth analysis of phishing contracts in real-world blockchains.  

\subsection{Overview of Detected Phishing Contracts}
We apply \codename to analyze smart contracts deployed up to August 2025 across four EVM-compatible blockchains: Ethereum mainnet, Binance Smart Chain (BSC), Avalanche (Avax), and Polygon. In total, we detected 4{,}224 phishing contracts, as summarized in Table~\ref{tab:overview}. Specifically, 480 phishing contracts were identified on Ethereum, 293 on BSC, 3{,}136 on Avax, and 315 on Polygon. On the Ethereum mainnet, among the 480 phishing contracts, 18 employed the \emph{storage-control} strategy, 453 adopted the \emph{Gas-control} strategy, and 9 used the \emph{timestamp-control} strategy. On BSC, 2 contracts were classified as \emph{storage-control}, 290 as \emph{Gas-control}, and 1 as \emph{timestamp-control}. On Avax, only 2 contracts leveraged the \emph{storage-control} strategy, while the remaining 3{,}134 contracts exclusively relied on \emph{Gas-control}. On Polygon, all 315 detected phishing contracts adopted the \emph{Gas-control} strategy.

\begin{table}[]
\caption{Overview of detected phishing contracts.}
\resizebox{\linewidth}{!}{%
\begin{tabular}{c|ccc|cc|c}
\hline
\multirow{2}{*}{\textbf{Blockchain}} & \multicolumn{3}{c|}{\textbf{Phishing Contract}}                                                                                                                                    & \multicolumn{2}{c|}{\textbf{Label Status}} & \multirow{2}{*}{\textbf{Total}} \\ \cline{2-6}
                                     & \begin{tabular}[c]{@{}c@{}}Storage\\ -control\end{tabular} & \begin{tabular}[c]{@{}c@{}}Gas-\\ control\end{tabular} & \begin{tabular}[c]{@{}c@{}}Timestamp\\ -control\end{tabular} & Yes                 & No                   &                                 \\ \hline
Ethereum                             & 18                                                         & 453                                                    & 9                                                            & 416                 & 64                   & 480                             \\
BSC                                  & 2                                                          & 290                                                    & 1                                                            & 257                 & 36                   & 293                             \\
Avax                                 & 2                                                          & 3,134                                                  & 0                                                            & 12                  & 3,124                & 3,136                           \\
Polygon                              & 0                                                          & 315                                                    & 0                                                            & 214                 & 101                  & 315                             \\ \hline
\textbf{Total}                       & 22                                                         & 4,192                                                  & 10                                                           & 899                 & 3,325                & 4,224                           \\ \hline
\end{tabular}
}
\label{tab:overview}
\end{table}

\textbf{Observations:} Overall, the distribution indicates that \emph{Gas-control} is the dominant phishing contract type, contributing more than 4,100 samples and accounting for 99.2\% of all detected phishing contracts. Among the four blockchains, Avalanche (Avax) is the most affected, with over 3,100 samples (74.2\%). Interestingly, our results show that \emph{External-control}, \emph{Gasprice-control}, and \emph{Blocknumber-control} phishing contracts have not been observed in the wild. We conjecture that the absence of \emph{External-control} contracts stems from their functional overlap with \emph{storage-control} contracts, while incurring higher implementation complexity and deployment costs. As a result, attackers may prefer the simpler and more cost-effective \emph{storage-control} contracts. For \emph{Gasprice-control} contracts, a likely explanation is the difficulty of selecting a stable and effective Gas price threshold. Most crypto wallets automatically set the gas price based on network congestion status, which can vary significantly over time. Although attackers could choose an extremely high threshold (e.g., 100K Gwei), doing so would require victims to hold a minimum balance of approximately 2.1 ETH\footnote{Assuming a minimum Gas consumption of 21{,}000 units}, thereby substantially reducing the pool of potential victims. The lack of \emph{Blocknumber-control} contracts may be attributed to the additional operational overhead required by phishing websites to obtain up-to-date block heights (e.g., via RPC calls). This introduces extra network latency and uncertainty, making it harder for attackers to reliably control the execution outcome of victims’ transactions. Therefore, these practical constraints likely discourage attackers from adopting \emph{Gasprice-control} and \emph{Blocknumber-control} strategies in real-world attacks.

\textbf{Labeling status:} All four blockchain explorers actively label addresses and contracts involved in phishing scams and other malicious activities based on off-chain intelligence and community or human reports, including Etherscan~\cite{me:etherscan} (Ethereum), BscScan~\cite{me:bscscan} (BSC), SnowScan~\cite{me:snowscan} (Avax), and PolygonScan~\cite{me:polygonscan} (Polygon). We query the labeling status of each detected phishing contract from the corresponding blockchain explorer to determine whether it has been assigned a ``phishing'' label. The results are summarized in Table~\ref{tab:overview}. Among the 480 phishing contracts detected on the Ethereum mainnet, 416 have been flagged as phishing by Etherscan, while the remaining 64 have not yet been labeled. On BSC, 257 contracts have been flagged, and 36 remain unlabeled. On Avax, only 12 phishing contracts have been flagged, whereas the other 3{,}124 contracts have not yet been labeled. On Polygon, 214 contracts have been flagged, while 101 remain unlabeled. Overall, these results suggest that although off-chain intelligence and human reports provide a useful mechanism for identifying phishing contracts, additional efforts are needed across all four blockchain explorers to comprehensively detect and label phishing contracts involved in transaction simulation phishing. This issue is particularly pronounced on Avax and Polygon, where a large fraction of the phishing contracts identified in our study remain unflagged.


\subsection{Timeline Trend of Phishing Contracts}
This section analyzes the timeline trend of phishing contracts. We retrieve each phishing contract's deployment date from the contract creation transaction and aggregate them by their deployment dates. 
\begin{figure}[htbp]
\hspace{-0.5cm}
    \centering
    \includegraphics[width=1.0\linewidth]{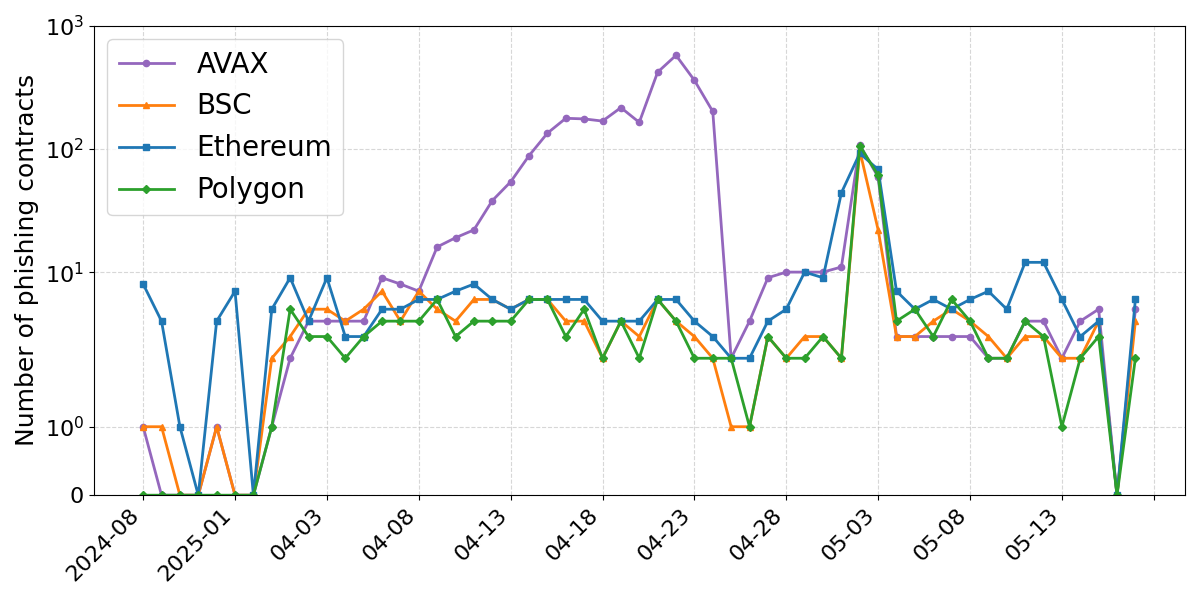}
    \caption{The timeline trend of phishing contracts deployed on four different blockchains.}
    \label{fig:contracts_timeline}
\end{figure}

\begin{table*}[]
\centering
\caption{Overview of victim transactions and the associated financial loss.}
\label{tab:victim}
\begin{tabular}{c|cccc|cc}
\hline
\multirow{2}{*}{\textbf{Blockchain}} & \multirow{2}{*}{\textbf{\begin{tabular}[c]{@{}c@{}}\# Lucky Tx\end{tabular}}} & \multirow{2}{*}{\textbf{\begin{tabular}[c]{@{}c@{}}\# Victim Tx\end{tabular}}} & \multirow{2}{*}{\textbf{\begin{tabular}[c]{@{}c@{}}\# Victim Address\end{tabular}}} & \multirow{2}{*}{\textbf{\begin{tabular}[c]{@{}c@{}}\# Profited Contract\end{tabular}}} & \multicolumn{2}{c}{\textbf{Financial Loss}} \\ \cline{6-7} 
                                     &                                                                                  &                                                                                   &                                                                                        &                                                                                         & Amount               & USD Value                 \\ \hline
Ethereum                             & 31                                                                               & 3,154                                                                             & 2,780                                                                                  & 257                                                                                        & 1,104.70 ETH          & \$3.19M
        \\
BSC                                  & 1                                                                                & 1,790                                                                             & 1,719                                                                                  & 168                                                                                        & 277.24 BNB           & \$182.32K           \\
Avax                                 & 0                                                                                & 215                                                                               & 213                                                                                    & 138                                                                                        & 4,308.33 AVAX         & \$111.53K         \\
Polygon                              & 114                                                                              & 1,064                                                                             & 1,030                                                                                  & 127                                                                                        & 9,602.15 POL          & \$2.15K              \\ \hline
Total                                & 146                                                                              & 6,223                                                                              & 5,742                                                                                  & 690                                                                                        & -                    & \$3.48M         \\ \hline
\end{tabular}
\end{table*}

Fig.~\ref{fig:contracts_timeline} shows the number of phishing contracts created on the four blockchains in different time periods, which reveals several interesting observations. First, the phishing contracts were introduced in Aug. 2024 on three blockchains: Ethereum mainnet, Avax, and BSC. On Polygon, the phishing contracts were introduced last, which occurred in March 2025. Second, before April 2025, the phishing contracts were occasionally created on the four blockchains, with fewer than 10 contracts deployed on several dates in the first seven months. Third, from April 8th to April 25th, 2025, the number of deployed phishing contracts on Avax experienced a significant growth, with the peak occurring on April 22nd, 2025, when there were more than 550 phishing contracts deployed in a single day. After that, the number dropped to below 10. During the same time period, the daily deployed phishing contracts on the other three blockchains remained at a much smaller number, which was less than 10. Then, later on May 3rd, 2025, the phishing contracts deployed on all four blockchains reached another peak, with more than 100 contracts deployed in a single day. Since then, the daily deployed phishing contracts on the four blockchains remained at a steady rate, mostly fluctuating between 5 and 11.

Overall, our timeline trend analysis indicates that the phishing contracts were first introduced by the attackers in Aug. 2024. Then, in the first seven months, the attackers were less active in deploying phishing contracts. Starting in Mar. 2025, the attackers became much more active and created $\approx$ 10 contracts daily, and the phishing contracts on three blockchains (Ethereum mainnet, BSC, and Polygon) follow a similar trend. However, on Avax, the attackers were much more active in deploying the phishing contracts. Then, on May 3rd, 2025, all four blockchains witnessed a spike of more than 100  phishing contracts deployed on the same day.

\subsection{Financial Impact of Phishing Contracts}
This section presents our analysis of the victim transactions and the associated financial loss. As previously mentioned, the success of the phishing contracts requires certain conditions to be met, such as the attacker's transaction successfully frontrunning the victim transactions and the victims manually modifying their transaction fields. Hence, the phishing contracts cannot achieve a 100\% success rate, and there could exist some lucky users who escaped from the phishing trap. We thereby define the lucky transaction and victim transaction as follows.
\begin{itemize}[leftmargin=*]
    \item \textbf{Lucky transaction:} If a transaction deposits funds to the phishing contract and then receives back more than the deposited amount, it is deemed a lucky transaction.
    \item \textbf{Victim transaction:} If a transaction deposits funds to the phishing contracts and the funds are immediately moved to another address, it is deemed a victim transaction.
\end{itemize}

\noindent After that, we then retrieve the transaction history of each phishing contract to determine the lucky and victim transactions. We first filter out transactions that were sent from the attackers themselves to avoid overestimating the scale of victims and the financial loss. We do so by checking each transaction's sender against the contract's deployer address and the funding address. The deployer address refers to the address sending the contract creation transaction to publish the phishing contract to the blockchain, and the funding address refers to the first address that deposits funds to the contract, serving as the contract’s initial bait for attracting victims. After removing such irrelevant transactions, we then determine the type of each remaining transaction based on the above criteria. For each victim transaction, we estimate the associated financial loss based on the market value of the transferred cryptocurrency assets at the time when the transaction was included in the blockchain. Specifically, for each blockchain, we referenced the historical price of its cryptocurrency asset from the corresponding explores (Etherscan, BscScan, SnowScan, and PolygonScan, respectively).

\begin{table*}[]
\caption{Profit distribution among different types of phishing contracts.}
\label{tab:dist_loss}
\begin{tabular}{c|cc|cc|cc|cc}
\hline
\multirow{2}{*}{\textbf{Blockchain}} & \multicolumn{2}{c|}{\textbf{Storage-control}} & \multicolumn{2}{c|}{\textbf{Gas-control}} & \multicolumn{2}{c|}{\textbf{Timestamp-control}} & \multicolumn{2}{c}{\textbf{Total}} \\ \cline{2-9} 
                                     & \# CA             & Profit (USD)            & \# CA          & Profit (USD)           & \# CA             & Profit (USD)              & \# CA     & Profit (\%)     \\ \hline
Ethereum                             & 12                & \$2.69M                     & 244            & \$186.59K                  & 1                 & \$305.71K                     & 257     & \$3.19M (91.50\%)         \\
BSC                                  & 2                 & \$168.26K                   & 166            & \$14.06K                   & -                 & -                           & 168       & \$182.32K (5.24\%)        \\
Avax                                 & 2                 & \$110.89K                   & 136            & \$639.16                   & -                 & -                           & 138       & \$111.53K (3.20\%)        \\
Polygon                              & -                 & -                           & 127            & \$2.15K                    & -                 & -                           & 127       & \$2.15K (0.06\%)          \\ \hline
\textbf{Total}                       & 16                & \$2.97M (85.37\%)           & 673            & \$203.44K (5.85\%)          & 1                & \$305.71K (8.78\%)             & 690    & \$3.48M                  \\ \hline
\end{tabular}
\end{table*}

\textbf{Overview of victim transactions:} Table~\ref{tab:victim} shows an overview of the victim transactions and the financial loss to the phishing contracts on the four blockchains. First of all, we can see that on the Ethereum mainnet, BSC, and Polygon, there are, respectively, 31, 1, and 114 lucky transactions that escaped from the phishing trap. On Avax, there is no lucky transaction. Although the lucky transactions have not lost funds, their profit from the phishing contracts is also negligible, as all the phishing contracts only transferred a tiny amount (10K wei) to them. Second, the table also shows that there is a total number of 6,223 victim transactions from 5,742 addresses on the four blockchains, indicating that some victim addresses lost funds twice or more to the phishing contracts. Among all the victim transactions, the Ethereum mainnet contributes the majority (3,154), and Avax contributes the lowest portion (215). In addition, it can be seen that the number of profited contract addresses (CA) on the four blockchains varies between 127 and 257, which is smaller than the number of deployed phishing contracts, indicating that not every phishing contract has made a profit. Among the four blockchains, phishing contracts on the Ethereum mainnet have the highest profitable rate, which is 257/480 = 53.5\%. Finally, the total financial loss from all the victim transactions reaches \$3.48 million USD. Among them, the Ethereum mainnet is still the dominant blockchain that contributes more than 1,104 ETH, which is worth \$3.19 million USD and accounts for 91.5\% of the total financial loss. On BSC, victims have lost more than 277 BNB, which is worth more than \$182K USD. On Avax and Polygon, victims have lost more than 4,300 AVAX and 9,600 POL, respectively worth \$111K USD and \$2.15K USD. Overall, our analysis suggests that the phishing attackers gained the majority of the profits on the Ethereum mainnet. One potential reason could be that compared to the other three cryptocurrencies, ETH has the highest value and is more popularly owned by users, which hence leads to more victim transactions and the majority of the financial loss on the Ethereum mainnet. In contrast, although Avax has the highest number of phishing contracts, due to most of them not making any profits, its victim transactions and financial loss thereby remained at a small scale.

 \begin{figure*}[htbp]
 \hspace{-0.4cm}
    \centering
    \subfloat[Trend of daily victim transactions.]{%
    \includegraphics[width=0.49\textwidth]{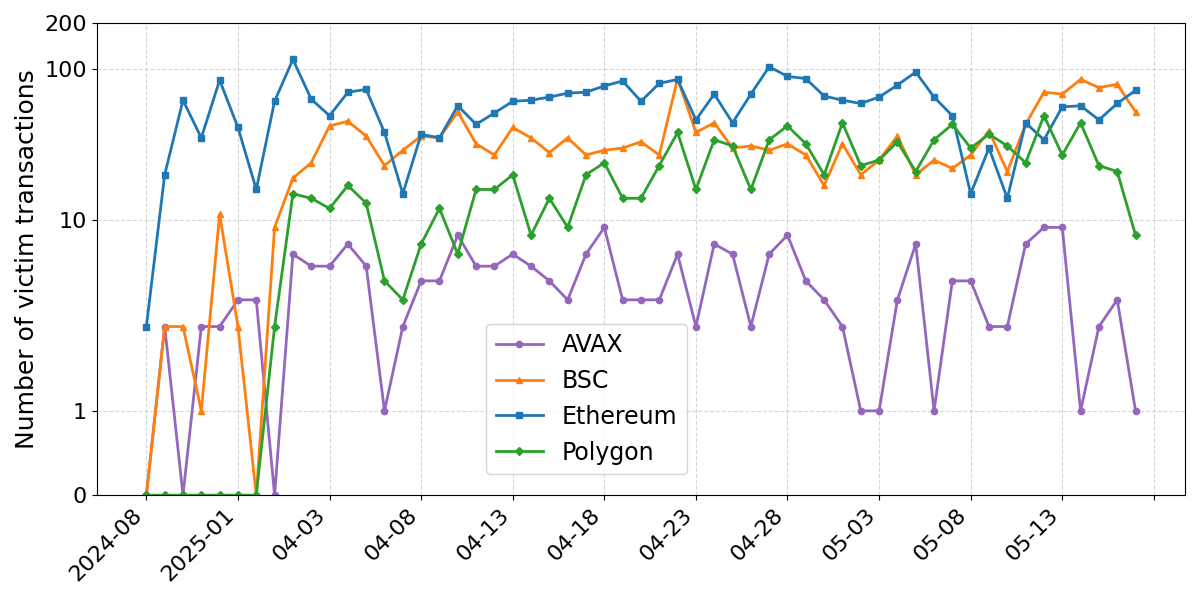}
    \label{fig:victim_tx_timeline}}
    \subfloat[Trend of daily financial loss in USD.]{%
    \includegraphics[width=0.49\textwidth]{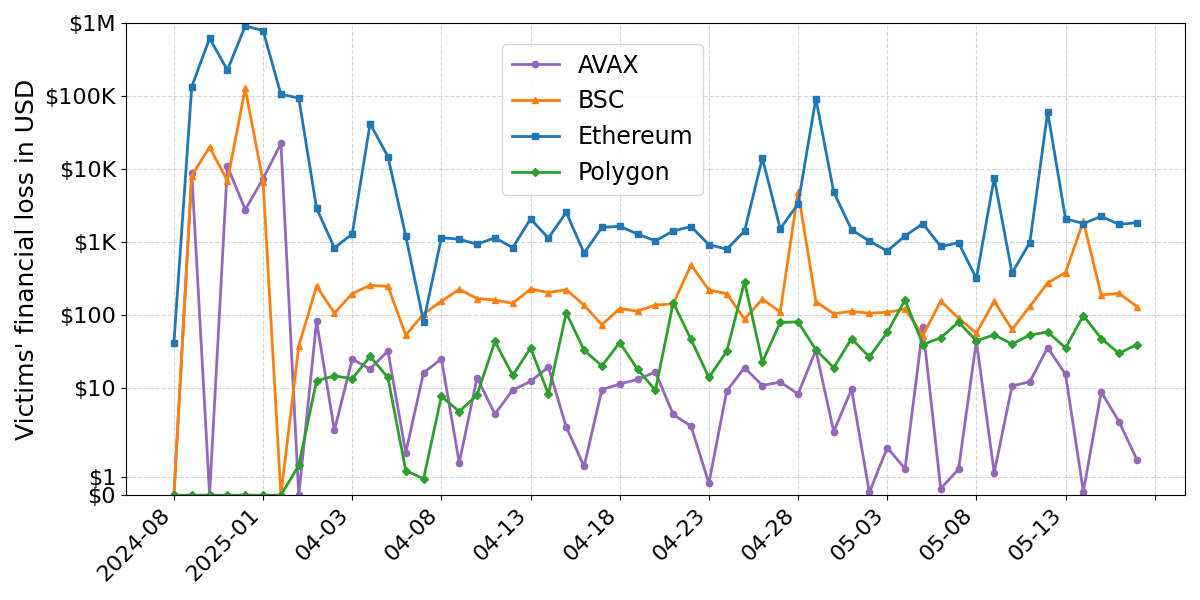}
    \label{fig:amount_timeline}}
    \caption{The timeline trend of victim transactions and the associated financial loss on the four blockchains.}
    \label{fig:trend_financial}
\end{figure*}

\textbf{Financial loss by phishing contract type:} In Table~\ref{tab:dist_loss}, we show the distribution of the profit collected by different types of phishing contracts on the four blockchains. From the table, we can obtain several interesting findings. First, it can be seen that the "storage-control" phishing contracts have gained most of the profits. With only 12 profited contracts on the Ethereum mainnet, the total profit gained by them exceeds \$2.69 million USD. A similar distribution can also be observed on BSC and Avax. With only 2 profited contracts on each blockchain, the total profit gained by them respectively exceeds \$168K and \$110K USD. In total, the profit collected by "storage-control" phishing contracts is above \$2.97 million USD, accounting for 85.3\% of our total uncovered profits. Second, compared to "storage-control", "Gas-control" has many more profited contracts, which vary between 127 and 244 on the four blockchains. However, the total profit collected by them is much lower, which is around \$203K USD. Third, the table also shows that "timestamp-control" has only one profited contract on the Ethereum mainnet. Despite that, the profit gained by the single contract exceeds \$305K USD. Overall, the distribution result suggests that "storage-control" is the most profitable type. We believe one potential reason is that the "storage-control" phishing contract can give the attacker a more deterministic control, as it does not rely on victims to modify certain transaction fields as required in the "Gas-control" contract, or rely on victims to submit transactions with some delays as required in the "timestamp-control" contract. The attacker of the "storage-control" contract can simply bid a high "Gasprice" in their transactions when adding the victim to the blacklist, which can ensure to frontrun the victims' transactions at a high success rate.

\textbf{Timeline trend of victims' financial loss:} We then present the timeline trend of victim transactions and the associated financial loss in Fig.~\ref{fig:trend_financial}. First of all, from Fig.~\ref{fig:victim_tx_timeline}, we can see that the earliest victim transaction occurred in Aug. 2024 on the Ethereum mainnet, BSC, and Avax, following a similar trend of the deployed phishing contracts. Since then, among the four blockchains, the Ethereum mainnet has always had the highest number of daily victim transactions until May 2025, varying mostly between 10 and 100. In the meantime, the daily number of victim transactions on BSC and Avax remains stable, mostly varying between 5 and 50. Then, starting May 9th, 2025, BSC surpasses the Ethereum mainnet to become the leader in the daily number of victim transactions. During the entire period, Polygon has the lowest number of victim transactions most of the time, which varies between 1 and 10. However, the timeline trend of the associated financial loss does not follow a similar pattern. As shown in Fig.~\ref{fig:amount_timeline}, the daily financial loss on the Ethereum mainnet is always one or more orders of magnitude higher than the other three blockchains, which is primarily due to the high value of ETH. In addition, the highest daily financial loss all occurred in Jan. 2025 on the three blockchains, including Ethereum mainnet, BSC, and Avax, respectively, with \$1 million USD, \$100K USD, and \$30K USD. After that, the daily financial loss on the four blockchains remains relatively stable, with occasional spikes on several dates such as Apr. 28th, 2025, and May 12th, 2025. Such a timeline trend indicates that phishing contracts usually can collect a much larger profit when they were first introduced. Later, when the blockchain and crypto wallet community has gained more knowledge, it becomes more difficult for them to sustain the same level of profit due to phishing contracts being continuously labelled and crypto wallet users becoming more cautious.


\textbf{Attackers' transaction cost:} We also look into the attackers' investment by measuring their transaction cost, which includes the transaction fee paid for deploying the phishing contracts, depositing initial funds to the phishing contracts, and adding users to the blacklist in the "storage-control" phishing contracts. The result is summarized in Table~\ref{tab:attack_cost}. We can see that on all four blockchains, the action of deploying the phishing contracts dominates the attackers' transaction cost, respectively costing 2.92 ETH, 1.04 BNB, 4.55 AVAX, and 46.25 POL. In comparison, depositing the initial funds to the phishing contracts only costs the attackers 0.08 ETH, 0.01 BNB, 0.04 AVAX, and 0.49 POL, respectively, on the four blockchains. Such an observation can be explained by that although deploying the phishing contracts and depositing the initial funds both incur one-time transaction cost, deploying phishing contracts is much more expensive action, with extra Gas paid for writing the contract bytecode to the blockchain storage, while depositing the initial funds does not involve such a cost and can be completed with the minimum transaction cost of 21,000 Gas. In addition, another part of the attackers' transaction cost comes from transactions blacklisting users in the "storage-control" contracts. On Ethereum, BSC, and Avax, the attackers respectively spent 21.75 ETH, 0.10 BNB, and 0.57 AVAX to blacklist users, while Polygon does not involve such a cost due to no "storage-control" contracts being detected. Adding up all the transaction costs, the attackers have respectively spent 24.75 ETH, 1.15 BNB, 5.16 AVAX, and 46.74 POL on four blockchains. Comparing with the profits on each blockchain shown in Table~\ref{tab:victim}, the return-on-investment (ROI) of this phishing attack is respectively 4,463\%, 24,108\%, 83,495\%, and 20,544\%.

\begin{table}[]
\caption{Overview of attackers' transaction cost and blacklisted users.}
\label{tab:attack_cost}
\resizebox{\linewidth}{!}{%
\begin{tabular}{c|cccc|c}
\hline
\multirow{2}{*}{\textbf{Blockchain}} & \multicolumn{4}{c|}{\textbf{Attackers' Transaction Cost}}                                                                                                                                                 & \multirow{2}{*}{\textbf{\begin{tabular}[c]{@{}c@{}}\# Blacklisted\\ Users\end{tabular}}} \\ \cline{2-5}
                                     & \begin{tabular}[c]{@{}c@{}}Contract \\ Deployment\end{tabular} & \begin{tabular}[c]{@{}c@{}}Initial \\ Funding\end{tabular} & \begin{tabular}[c]{@{}c@{}}Blacklisting \\ Users\end{tabular} & Subtotal (USD) &                                                                                          \\ \hline
Ethereum                             & 2.92 ETH                                                       & 0.08 ETH                                                   & 21.75 ETH                                                     & 24.75 ETH (\$77.3K)   & 3,160                                                                                    \\
BSC                                  & 1.04 BNB                                                       & 0.01 BNB                                                   & 0.10 BNB                                                      & 1.15 BNB (\$690)    & 208                                                                                      \\
Avax                                 & 4.55 AVAX                                                      & 0.04 AVAX                                                  & 0.57 AVAX                                                     & 5.16 AVAX (\$121)  & 124                                                                                      \\
Polygon                              & 46.25 POL                                                      & 0.49 POL                                                   & -                                                             & 46.74 POL (\$10)   & -                                                                                        \\ \hline
\end{tabular}
}
\end{table}
             
\textbf{Blacklisted users in "storage-control":} Among the phishing contracts, one unique characteristic of the "storage-control" phishing contract is that, in order to victimize users, the attackers have to send transactions to add users to the blacklist. Thanks to this characteristic, we can thereby trace the phishing contract's transaction history to estimate the number of users that have been blacklisted in the "storage-control" phishing contracts. We do so by looking for transactions that add users to the blacklist (e.g., those invoking the \texttt{add()} function) and then retrieving the blacklisted address from the transaction's input field. The result is presented in Table~\ref{tab:attack_cost}. We can see that there are, respectively, 3,160, 208, 124 unique addresses blacklisted in the "storage-control" phishing contracts on the Ethereum mainnet, BSC, and Avax. Such a result reveals the potential number of users who have visited the phishing website established by the attacker. In addition, we can also see that the number of blacklisted addresses is higher than the actual number of victim addresses on the three blockchains, implying that there were some cautious users who visited the attacker's phishing website but did not fall victim.
 
\subsection{Phishing Contract Clusters}
The phishing contracts detected in our work are not operated independently but rather have connections with each other, which can thus form clusters. In this section, we cluster the phishing contracts with the following criteria: (1) If two phishing contracts are deployed by the same blockchain address, we group them together; (2) If the initial funds of two phishing contracts are deposited from the same address, we group them together; (3) If the victim's funds sent to two phishing contracts are internally transferred to the same address, we group them together.

Applying the above criteria to the detected phishing contracts, we identify 8 clusters comprising 724 contracts in total. Among them, one dominant cluster emerges on each blockchain and accounts for the majority of the profits on that chain, as summarized in Table~\ref{tab:cluster}. In addition, we identify four small clusters, with sizes ranging from 2 to 8 contracts. Three of these appear on the Ethereum mainnet and one on Avax. However, we omit them from the table because none of them generated measurable profit. As shown in Table~\ref{tab:cluster}, the largest cluster on the Ethereum mainnet controls 268 phishing contracts and has extracted more than \$2.88 million USD, accounting for 90.4\% of the profits on Ethereum (approximately 83\% of the total profits across all blockchains). The remaining three major clusters are found on BSC, Avax, and Polygon, respectively. Each controls between 130 and 169 contracts and captures 100\% of the total profit observed on its corresponding blockchain.

\begin{table}[]
\caption{Top phishing contract clusters and their profits.}
\label{tab:cluster}
\resizebox{\linewidth}{!}{%
\begin{tabular}{c|c|c|c|c}
\hline
\textbf{Cluster} & \textbf{\# Contract} & \textbf{Profit in USD (\%)} & \textbf{Blockchain} & \textbf{Most Profited Contract}             \\ \hline
1                   & 268                  & \$2.88M (90.4\%)              & Ethereum            & 0x4805a2***108433 \\ \hline
2                   & 169                  & \$182.32K (100\%)             & BSC                 & 0xcada58***5c3b78 \\ \hline
3                   & 142                  & \$111.53K (100\%)             & Avax                & 0x00000c***d00000 \\ \hline
4                   & 130                  & \$2.15K (100\%)               & Polygon             & 0x000049***960000 \\ \hline
\end{tabular}
}
\end{table}

\subsection{Case Study of Top Profited Contracts}
We first show the top 10 most profitable phishing contracts in Table~\ref{tab:top10}, ranked by their profits in USD. We have confirmed that they all belong to phishing contracts through verifying their transaction behaviors and money flow. As can be seen from the table, 9 contracts are deployed on the Ethereum mainnet, and only one contract (ranked at the 5th) is deployed on BSC. In addition, among the 10 contracts, 9 contracts employed the "storage-control" strategy, and only one (ranked at the 3rd) employed the "timestamp-control" strategy. The most profited phishing contract is deployed on the Ethereum mainnet, which solely profited more than \$1.2 million USD from 164 victim transactions through the "storage-control" strategy. The second most profited contract also came from the Ethereum mainnet and utilized the "storage-control" strategy. However, this contract profited more than \$477K USD in a single victim transaction. The 3rd and 4th most profited contracts are also deployed on the Ethereum mainnet, which respectively profited more than \$305K and \$290K USD, and both contributed 13 victim transactions. The 5th most profited contract is deployed on BSC, which profited over \$151K USD from 14 victim transactions. The last 5 phishing contracts (6th to 10th) are all deployed on the Ethereum mainnet and employ the "storage-control" strategy. Their profits range from \$91K to \$140K USD, with the number of victim transactions varying between 8 and 22. The profit collected by the 10 contracts reaches \$3 million USD, accounting for more than 86\% of the total profit uncovered in our work.

\begin{table}[]
\caption{The top 10 most profited contracts ranked by USD.}
\label{tab:top10}
\resizebox{\linewidth}{!}{%
\begin{tabular}{c|c|c|c|c|c}
\hline
\textbf{Rank} & \textbf{Contract Address}                           & \textbf{Phishing Type}     & \# \textbf{Victim Tx} & \textbf{Profit (USD)} & \textbf{Blockchain} \\ \hline
1  & 0x4805a2***108433 & storage-control   & 164          & \$1,206,057.38       & Ethereum   \\ \hline
2  & 0x000008***700000 & storage-control   & 1            & \$477,195.48         & Ethereum   \\ \hline
3  & 0xfef6b0***8270e5 & timestamp-control & 13           & \$305,707.27         & Ethereum   \\ \hline
4  & 0x00000c***d00000 & storage-control   & 13           & \$290,666.14         & Ethereum   \\ \hline
5  & 0xcada58***5c3b78 & storage-control   & 14           & \$151,644.47         & BSC        \\ \hline
6  & 0x00000c***200000 & storage-control   & 22           & \$140,098.19         & Ethereum   \\ \hline
7  & 0x478e1c***6a6825 & storage-control   & 8            & \$127,076.65         & Ethereum   \\ \hline
8  & 0x00000c***e00000 & storage-control   & 14           & \$106,578.39         & Ethereum   \\ \hline
9  & 0x000007***f00000 & storage-control   & 12           & \$104,530.76         & Ethereum   \\ \hline
10 & 0x000000***e00000 & storage-control   & 8            & \$91,487.93          & Ethereum   \\ \hline
\end{tabular}
}
\end{table}

\textbf{Most profited phishing contract:} The most profited phishing contract is deployed on the Ethereum mainnet at address \hash{0x4805a2***108433}. The contract was created by address \hash{0x220DC8***C945C6} in transaction \hash{0xb5cc9c9b***c72a0c83} on Aug. 27th, 2024, at block height 20,619,264. On the same day, right after the creation, the attacker sent two transactions to invoke the \texttt{Add()} function to add two targeted users to the blacklist. However, none of the targeted users were successfully deceived because there were no initial funds in the contract. Due to this, users cannot see the profitable opportunity in their transaction simulation results. The attacker quickly realized the problem and then sent four transactions to transfer a tiny amount of ETH (100K wei) to the contract on the same day. After that, another user was targeted and added to the blacklist by the attacker on block \# 20,620,183. In the same block, the targeted user sent a transaction to invoke the \texttt{Claim()} function and transferred more than 0.009 ETH. The user's funds were immediately moved to address \hash{0x000037***Ce0000}, making the user fall victim. After 41 blocks, another victim was added to the blacklist and then lost more than 0.007 ETH in the next block. Since then, the attacker kept adding targeted users to the blacklist. Occasionally, there were some users who fell victim. The largest financial loss to this phishing contract was from the victim address \hash{0x116bd2***c546ad}, which was added to the blacklist at block \# 21323413. Then, the victim transferred more than 18.4 ETH to the phishing contract at block \# 21323426.

\textbf{Victim with the largest loss:} Among all victim addresses identified in our study, the largest single-transaction loss was from address \hash{0x98Ddab***cD84C2}, which transferred more than 143.4 ETH to a \emph{storage-control} phishing contract. The phishing contract, deployed at address \hash{0x000008***700000}, ranks second in Table~\ref{tab:top10} despite profiting from only a single victim transaction. The contract was deployed by address \hash{0x505506***77BBa9} on Jan. 8, 2025, at block \#21579459. Shortly thereafter, at block \#21579469, the same address funded the contract with 0.028 ETH as initial funds. Following deployment, the attacker issued multiple transactions to add addresses to the contract's blacklist. Later that day, at block \#21581912, the victim's address was added to the blacklist. Three blocks later, the victim invoked the \texttt{claim()} function, transferring 143.4 ETH to the phishing contract. The transferred funds were immediately redirected to the attacker’s profit-collection address \hash{0x000037***Ce0000} via an internal call.

\begin{figure}[!htbp]
\hspace{-0.5cm}
     \centering
     \includegraphics[width=0.90\linewidth]{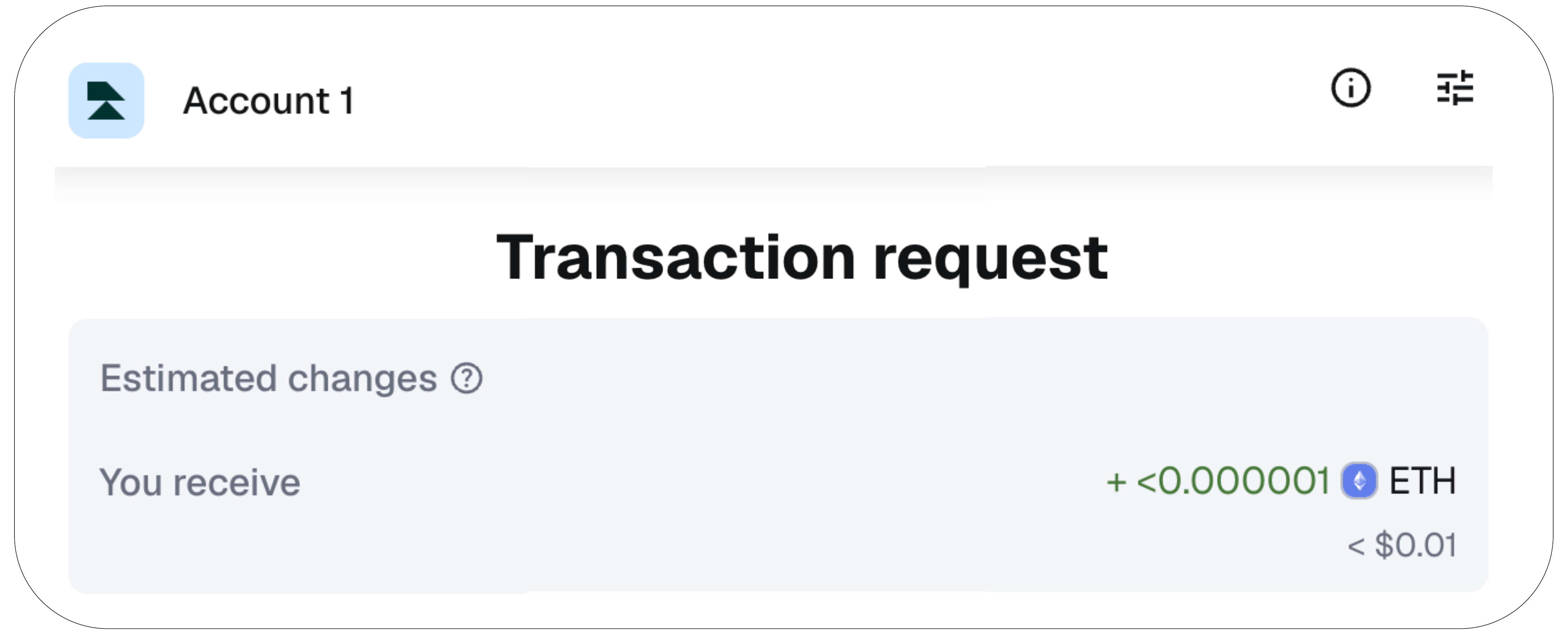}
     \caption{The transaction simulation result displayed on MetaMask where the phishing contract returns 1 wei.}
     \label{fig:sim}
\end{figure}
\section{Discussion}
\label{sec:discuss}
This section discusses design flaws in crypto wallets when presenting transaction simulation results and recommends potential countermeasures to mitigate this phishing threat. We then outline several limitations of our work.

\subsection{Widespread Inaccurate Simulations}
In our case study, we show that a victim lost more than 143 ETH to a \emph{storage-control} phishing contract. However, based on our runtime verification, the reward amount returned by the contract is only 1 wei (\num{e-18} ETH). In fact, the typical returned reward by the phishing contracts detected in our study ranges between 1 and 10,000 wei, a tiny and negligible amount. A natural question arises: why would victims approve transactions that transfer a large portion or even all of their assets in exchange for such a trivial reward?

We hypothesize that misleading transaction simulation results presented by crypto wallets play a key role in deceiving users into approving these transactions. To verify this hypothesis, we analyze 11 leading crypto wallets that provide transaction simulation features, including MetaMask, Rabby, and Phantom. For each wallet, we initiate a transaction that transfers our entire account balance (e.g., 0.5 ETH) to a phishing contract that returns 1 wei, and then examine the transaction simulation preview displayed on the wallet’s UI. Surprisingly, we find that most wallets present misleading or inaccurate simulation results. For example, as shown in Fig.~\ref{fig:sim}, MetaMask indicates that executing the transaction would increase our account balance by an amount smaller than 0.000001 ETH. Such a result can easily be misinterpreted by users as a net gain close to 0.000001 ETH. In reality, the returned amount (1 or even 10,000 wei) is far from sufficient to cover the transaction cost, meaning that the user's actual post-transaction balance change should be negative rather than positive. Similar misleading and inaccurate simulation issues are also observed in other wallets (details are provided in Appendix~\ref{sec:screenshots}). Beyond misleading and inaccurate balance changes, we identify another design issue: most wallets display only the predicted post-transaction balance change in the simulation preview, rather than explicitly showing the exact amount being transferred. As a result, users are not informed that the transaction will transfer their entire account balance to the contract. This lack of visibility into the transferred amount severely limits users’ ability to assess transaction risk.

In summary, we believe that misleading and inaccurate simulation results, together with the lack of transparency regarding transferred amounts, jointly contribute to victims approving malicious transactions and suffering substantial financial losses.


\subsection{Recommended Countermeasures}


To effectively mitigate transaction simulation–based phishing threats, the key is to ensure that wallets present accurate and up-to-date simulation results. Based on our analysis, we recommend the following countermeasures for each type of phishing contract:

\begin{itemize}[leftmargin=*]
\item \textbf{Storage-control/External-control}: Since this type of phishing contract relies on manipulating contract state, the transaction simulation service should monitor whether the state of the interacted contract, as well as any internally invoked contracts, has changed since the last simulation. If any state change is detected, the transaction should be re-simulated, and the updated results should be presented to the user.


\item \textbf{Gas-control}: The transaction simulation service should avoid using a default \texttt{gaslimit} and instead simulate the transaction using the \texttt{gaslimit} specified in the transaction request. Furthermore, if the user modifies the \texttt{gaslimit}, the transaction should be re-simulated before signing.

\item \textbf{Gasprice-control}: Similarly, the transaction simulation service should simulate execution using the transaction’s specified \texttt{gasprice}, and re-simulate the transaction whenever the user modifies this parameter.

\item \textbf{Blocknumber-control}: As this phishing contract leverages \texttt{block.number} to alter control flow, the simulation service should execute the transaction under both the current block number and a future block number to detect any divergence in execution outcomes.


\item \textbf{Timestamp-control}: Likewise, the simulation service should simulate the transaction using both the current block timestamp and a future timestamp to identify discrepancies caused by time-dependent control logic.

\end{itemize}

\noindent Besides, we recommend that wallets also integrate an recipient verification service to protect users, as done in popular wallets such as Coinbase~\cite{me:coinbase}. This service raises warnings when users transfer to smart contracts flagged on blockchain explorers. 

\subsection{Limitations}
Despite detecting more than 4,000 phishing contracts on real-world blockchains, our work has several limitations. First, although we systematically analyzed dynamic variables in the blockchain execution environment to construct our phishing contract taxonomy, attackers may combine multiple variables or exploit alternative strategies that are not covered in this study. For example, phishing contracts could leverage other mutable on-chain states, such as token exchange rates, to control transfer logic. However, we believe such strategies are complex to implement and difficult to deterministically control the execution outcome. Second, \codename has limited effectiveness against heavily obfuscated phishing contracts. While such cases are rare in our evaluation, advanced obfuscation techniques can hinder static analysis and symbolic execution. This limitation could potentially be mitigated by incorporating de-obfuscation techniques~\cite{yang2025insecurity} or relying more extensively on runtime verification, albeit at the cost of increased execution and analysis overhead. Third, although we carefully filtered out transactions initiated by attackers when quantifying financial losses, mis-classification remains possible. Specifically, if attackers interacted with phishing contracts using addresses not directly associated with the phishing activity (e.g., for testing), these transactions may be mistakenly counted as victim losses. Consequently, our loss estimates should be interpreted as an upper bound rather than exact ground truth. Similarly, our clustering analysis relies solely on observable on-chain associations. Phishing contracts controlled by the same entity may remain unlinked if no such associations are visible, potentially resulting in an overestimation of the number of attacker clusters. Finally, the reported return-on-investment (ROI) accounts only for on-chain transaction costs, including contract deployment, initial funding, and blacklist updates. Other costs, such as domain registration, phishing website promotion, and fund laundering, are difficult to quantify due to limited visibility and are therefore excluded from our analysis.

\section{Related Work}
\label{sec:related}
Existing work has studied various cryptocurrency phishings and scams, including Ponzi Schemes~\cite{kell2021forsage, bian2021image, xia20covidscams, bartoletti2020dissecting, bartoletti2018data, chen2018detecting}, fraudulent Initial Coin Offering~\cite{phua2022don, chiu2022using, liebau2019crypto, zetzsche2017ico}, fake exchange scams~\cite{xia2020characterizing}, phishing scams~\cite{chen2020phishing, badawi2020automatic}, giveaway scams~\cite{xia20covidscams, vakilinia2022cryptocurrency, xigao2023doublenothing, li2023understanding}, honeypot contract scams~\cite{torres2019art, chen2020honeypot}, fake token scam~\cite{gao2020tracking, xia21scams}, arbitrage bot phishing~\cite{li2023towards}, and blockchain address poisoning~\cite{guan24ccs,tsuchiya2025blockchainaddresspoisoning,chen2025dissecting}.

Some of the existing works also proposed new approaches to detect the involved phishing and scam activities. For instance, Li et al.~\cite{xigao2023doublenothing} developed a free giveaway scam detection system to collect suspicious registered domains from Certificate Transparency Log (CTLog) and discovered more than 10K giveaway scam URLs. Xia et al.~\cite{xia20covidscams} developed a scam detection system for cryptocurrency scams using keywords such as "COVID-19" and "cryptocurrency" on social media platforms, including Twitter, Telegram, Etherscan, etc. Xia et al.~\cite{xia21scams} developed a scam token detection system based on machine learning models to systematically analyze all tokens listed on the popular DEX, Uniswap. Their work identified over 10K scam tokens that collected \$16 million USD. Li et al.~\cite{li2023towards} developed a detection system to identify phishing arbitrage bot contracts that directly transfer funds from the victims' addresses to the attackers' addresses. Their work detected more than 20K phishing contracts on Ethereum mainnet and BSC, and quantified that more than 25K victims lost $\approx$\$15 million USD. Most recently, several studies~\cite{guan24ccs,tsuchiya2025blockchainaddresspoisoning, chen2025dissecting} have developed a detection system to detect phishing transfers involved in the address poisoning attack. These works quantified that the victim's total financial loss was up to \$100 million USD. He et al.~\cite{he2023txphishscope} developed a detection system for transaction-based phishing attacks launched on fake websites that directly transfer funds from the victim to the attacker. Their detection system works by monitoring Certificate Transparency Log~\cite{ctlog} to identify suspicious domains and then visiting the suspicious website to trigger the transaction signing operations to detect phishing addresses.


\section{Conclusion}
\label{sec:conclude}
In this work, we investigate \emph{transaction simulation phishing} and present the first comprehensive study of this emerging threat. We develop a taxonomy of phishing contracts and design \codename, a bytecode-level detection system that combines static and dynamic analysis to identify phishing contracts. Applying \codename to the Ethereum mainnet, Binance Smart Chain, Avalanche, and Polygon, we detect over 4,000 phishing contracts and uncover more than 5,700 victims who collectively lost approximately \$3.48 million USD. These findings highlight the severity of transaction simulation phishing. We hope this work raises awareness of this previously underexplored attack vector and motivates the development of more robust transaction simulation mechanisms and effective countermeasures across the ecosystem.


\section{Generative AI Usage}
This paper was edited for grammar using ChatGPT. In addition, due to the lack of a ground-truth dataset for phishing contracts, Gemini is used to produce the initial 30 phishing contracts, which are used to construct the ground-truth dataset. The authors have validated the contract code generated by Gemini by manual verification as well as experiment runs to ensure its quality, functionality, and accuracy.

\bibliographystyle{plain}
\bibliography{shixuan.bib, bkc.bib, bkcscam.bib, twitterscam.bib, youtubescam.bib}

@misc{ctlog,
title = {Crt.sh},
howpublished = {\url{https://certificate.transparency.dev}}
}

@misc{me:blockprop,
title = {Block and Transaction Properties},
howpublished = {\url{https://docs.soliditylang.org/en/latest/units-and-global-variables.html#block-and-transaction-properties}}
}

@misc{loss2025,
title = {2025 Crypto Crime Mid-year Update: Stolen Funds Surge as DPRK Sets New Records},
howpublished = {\url{https://www.chainalysis.com/blog/2025-crypto-crime-mid-year-update}}
}

@misc{me:similarAPI,
title = {Similar Contracts Search},
howpublished = {\url{https://etherscan.io/find-similar-contracts}}
}

@misc{me:etherscan,
title = {Etherscan: Ethereum (ETH) Blockchain Explorer},
howpublished = {\url{https://etherscan.io/}}
}

@misc{me:gemini,
title = {Google Gemini},
howpublished = {\url{https://gemini.google.com/}}
}

@misc{me:greedsolver,
title = {Reachibility of a CALL statement},
howpublished = {\url{https://ucsb-seclab.github.io/greed/examples/#1-reachibility-of-a-call-statement}}
}

@misc{me:fortaset,
title = {A dataset of malicious and benign smart contracts},
howpublished = {\url{https://huggingface.co/datasets/forta/malicious-smart-contract-dataset}}
}

@misc{me:fortaml,
title = {How Forta’s Predictive ML Models Detect Attacks Before Exploitation},
howpublished = {\url{https://forta.org/blog/how-fortas-predictive-ml-models-detect-attacks-before-exploitation}}
}

@misc{me:slithersm,
title = {Code Similarity: slither-simil},
howpublished = {\url{https://secure-contracts.com/program-analysis/slither/docs/src/tools/Code-Similarity-Detector.html}}
}

@misc{me:tac,
title = {Three-address code},
howpublished = {\url{https://en.wikipedia.org/wiki/Three-address_code}}
}

@misc{me:fordefi,
title = {FORDEFI - Simulate Transactions},
howpublished = {\url{https://docs.fordefi.com/developers/simulate-transactions#simulation-data
}}
}

@misc{me:etl,
title = {Ethereum ETL},
howpublished = {\url{https://github.com/blockchain-etl/ethereum-etl}}
}

@misc{me:foundry,
title = {foundry - Ethereum Development Framework},
howpublished = {\url{https://getfoundry.sh}}
}

@misc{me:bscscan,
title = {BscScan: BNB Smart Chain Explorer},
howpublished = {\url{https://bscscan.com/
}}
}

@misc{me:snowscan,
title = {Avalanche C-Chain (AVAX) Blockchain Explorer},
howpublished = {\url{https://snowscan.xyz}}
}

@misc{me:polygonscan,
title = {Polygon PoS Chain Explorer},
howpublished = {\url{https://polygonscan.com}}
}

@misc{me:evm,
   title = {An Ethereum Virtual Machine Opcodes Interactive Reference},
   howpublished = {\url{https://www.evm.codes}}
}

@inproceedings{xigao2023doublenothing,
    title     = {Double and Nothing: Understanding and Detecting Cryptocurrency Giveaway Scams},
    author    = {Li, Xigao and Yepuri, Anurag and Nikiforakis, Nick},
    journal = {Network and Distributed Systems Security (NDSS) Symposium},
    year      = {2023}
}

@misc{me:sol,
   title = {Solidity programming language},
   howpublished = {\url{https://solidity.readthedocs.io/en/develop/}}
}

@misc{me:knownattacks,
   title = {Known Attacks - Ethereum Smart-Contract Best Practices},
   howpublished = {\url{https://consensysdiligence.github.io/smart-contract-best-practices}}
}

@inproceedings{ChengHLZLLR19,
  title={Towards a first step to understand the cryptocurrency stealing attack on ethereum},
  author={Cheng, Zhen and Hou, Xinrui and Li, Runhuai and Zhou, Yajin and Luo, Xiapu and Li, Jinku and Ren, Kui},
  booktitle={22nd international symposium on research in attacks, intrusions and defenses (RAID 2019)},
  pages={47--60},
  year={2019}
}

@misc{me:metamask,
 title = {MetaMask: The Ultimate Crypto Wallet for DeFi, Web3 Apps}, 
 howpublished = {\url{https://metamask.io/}}
}

@inproceedings{meisami2025sigscope,
  title={SigScope: Detecting and Understanding Off-Chain Message Signing-related Vulnerabilities in Decentralized Applications},
  author={Meisami, Sajad and Dabadie, Hugo and Li, Song and Tang, Yuzhe and Duan, Yue},
  booktitle={Proceedings of the ACM on Web Conference 2025},
  pages={4284--4299},
  year={2025}
}

@inproceedings{yan2024stealing,
  title={Stealing trust: Unraveling blind message attacks in web3 authentication},
  author={Yan, Kailun and Zhang, Xiaokuan and Diao, Wenrui},
  booktitle={Proceedings of the 2024 on ACM SIGSAC Conference on Computer and Communications Security},
  pages={555--569},
  year={2024}
}

@inproceedings{abdelaziz2023smart,
  title={Smart learning to find dumb contracts},
  author={Abdelaziz, Tamer and Hobor, Aquinas},
  booktitle={32nd USENIX Security Symposium (USENIX Security 23)},
  pages={1775--1792},
  year={2023}
}

@inproceedings{gritti2023confusum,
title={Confusum contractum: confused deputy vulnerabilities in ethereum smart contracts},
author={Gritti, Fabio and Ruaro, Nicola and McLaughlin, Robert and Bose, Priyanka and Das, Dipanjan and Grishchenko, Ilya and Kruegel, Christopher and Vigna, Giovanni},
booktitle={32nd USENIX Security Symposium (USENIX Security 23)},
pages={1793--1810},
year={2023}
}

@inproceedings{ruaro2024crush,
title={Not your Type! Detecting Storage Collision Vulnerabilities in Ethereum Smart Contracts},
author={Ruaro, Nicola and Gritti, Fabio and McLaughlin, Robert and Grishchenko, Ilya and Kruegel, Christopher and Vigna, Giovanni},
booktitle={Network and Distributed Systems Security (NDSS) Symposium 2024},
year={2024}
}

@article{he2025phishing,
  title={Phishing Tactics Are Evolving: An Empirical Study of Phishing Contracts on Ethereum},
  author={He, Bowen and Hu, Xiaohui and Hu, Yufeng and Yu, Ting and Chang, Rui and Wu, Lei and Zhou, Yajin},
  journal={Proceedings of the ACM on Measurement and Analysis of Computing Systems},
  volume={9},
  number={2},
  pages={1--24},
  year={2025},
  publisher={ACM New York, NY, USA}
}

@article{yang2025insecurity,
  title={Insecurity Through Obscurity: Veiled Vulnerabilities in Closed-Source Contracts},
  author={Yang, Sen and Qin, Kaihua and Yaish, Aviv and Zhang, Fan},
  journal={arXiv preprint arXiv:2504.13398},
  year={2025}
}

@article{kell2021forsage,
  title={Forsage: Anatomy of a smart-contract pyramid scheme},
  author={Kell, Tyler and Yousaf, Haaroon and Allen, Sarah and Meiklejohn, Sarah and Juels, Ari},
  journal={arXiv preprint arXiv:2105.04380},
  year={2021}
}

@article{bartoletti2020dissecting,
  title={Dissecting Ponzi schemes on Ethereum: identification, analysis, and impact},
  author={Bartoletti, Massimo and Carta, Salvatore and Cimoli, Tiziana and Saia, Roberto},
  journal={Future Generation Computer Systems},
  volume={102},
  pages={259--277},
  year={2020},
  publisher={Elsevier}
}

@inproceedings{bartoletti2018data,
  title={Data mining for detecting bitcoin ponzi schemes},
  author={Bartoletti, Massimo and Pes, Barbara and Serusi, Sergio},
  booktitle={2018 Crypto Valley Conference on Blockchain Technology (CVCBT)},
  pages={75--84},
  year={2018},
  organization={IEEE}
}

@article{bian2021image,
  title={Image-based scam detection method using an attention capsule network},
  author={Bian, Lingyu and Zhang, Linlin and Zhao, Kai and Wang, Hao and Gong, Shengjia},
  journal={IEEE Access},
  volume={9},
  pages={33654--33665},
  year={2021},
  publisher={IEEE}
}

@inproceedings{chen2018detecting,
  title={Detecting ponzi schemes on ethereum: Towards healthier blockchain technology},
  author={Chen, Weili and Zheng, Zibin and Cui, Jiahui and Ngai, Edith and Zheng, Peilin and Zhou, Yuren},
  booktitle={Proceedings of the 2018 world wide web conference},
  pages={1409--1418},
  year={2018}
}

@article{phua2022don,
  title={Don't trust, verify: The economics of scams in initial coin offerings},
  author={Phua, Kenny and Sang, Bo and Wei, Chishen and Yu, Gloria Yang},
  journal={Available at SSRN 4064453},
  year={2022}
}

@article{chiu2022using,
  title={Using textual analysis to detect initial coin offering frauds},
  author={Chiu, Tiffany and Chiu, Victoria and Wang, Tawei and Wang, Yunsen},
  journal={Journal of Forensic Accounting Research},
  volume={7},
  number={1},
  pages={165--183},
  year={2022},
  publisher={American Accounting Association}
}

@article{liebau2019crypto,
  title={Crypto-currencies and icos: Are they scams? an empirical study},
  author={Liebau, Daniel and Schueffel, Patrick},
  journal={An Empirical Study (January 23, 2019)},
  year={2019}
}

@article{zetzsche2017ico,
  title={The ICO Gold Rush: It's a scam, it's a bubble, it's a super challenge for regulators},
  author={Zetzsche, Dirk A and Buckley, Ross P and Arner, Douglas W and F{\"o}hr, Linus},
  journal={University of Luxembourg Law Working Paper},
  number={11},
  pages={17--83},
  year={2017}
}

@inproceedings{chen2020phishing,
  title={Phishing Scam Detection on Ethereum: Towards Financial Security for Blockchain Ecosystem.},
  author={Chen, Weili and Guo, Xiongfeng and Chen, Zhiguang and Zheng, Zibin and Lu, Yutong},
  booktitle={IJCAI},
  volume={7},
  pages={4456--4462},
  year={2020}
}

@inproceedings{badawi2020automatic,
  title={An automatic detection and analysis of the bitcoin generator scam},
  author={Badawi, Emad and Jourdan, Guy-Vincent and Bochmann, Gregor and Onut, Iosif-Viorel},
  booktitle={2020 IEEE European Symposium on Security and Privacy Workshops (EuroS\&PW)},
  pages={407--416},
  year={2020},
  organization={IEEE}
}

@article{gao2020tracking,
  title={Tracking counterfeit cryptocurrency end-to-end},
  author={Gao, Bingyu and Wang, Haoyu and Xia, Pengcheng and Wu, Siwei and Zhou, Yajin and Luo, Xiapu and Tyson, Gareth},
  journal={Proceedings of the ACM on Measurement and Analysis of Computing Systems},
  volume={4},
  number={3},
  pages={1--28},
  year={2020},
  publisher={ACM New York, NY, USA}
}

@inproceedings{vakilinia2022cryptocurrency,
  title={Cryptocurrency Giveaway Scam with YouTube Live Stream},
  author={Vakilinia, Iman},
  booktitle={2022 IEEE 13th Annual Ubiquitous Computing, Electronics \& Mobile Communication Conference (UEMCON)},
  pages={0195--0200},
  year={2022},
  organization={IEEE}
}

@article{xia2020characterizing,
  title={Characterizing cryptocurrency exchange scams},
  author={Xia, Pengcheng and Wang, Haoyu and Zhang, Bowen and Ji, Ru and Gao, Bingyu and Wu, Lei and Luo, Xiapu and Xu, Guoai},
  journal={Computers \& Security},
  volume={98},
  pages={101993},
  year={2020},
  publisher={Elsevier}
}

@article{xia21scams,
author = {Xia, Pengcheng and Wang, Haoyu and Gao, Bingyu and Su, Weihang and Yu, Zhou and Luo, Xiapu and Zhang, Chao and Xiao, Xusheng and Xu, Guoai},
title = {Trade or Trick? Detecting and Characterizing Scam Tokens on Uniswap Decentralized Exchange},
year = {2021},
issue_date = {December 2021},
publisher = {Association for Computing Machinery},
address = {New York, NY, USA},
volume = {5},
number = {3},
url = {https://doi.org/10.1145/3491051},
doi = {10.1145/3491051},
journal = {Proc. ACM Meas. Anal. Comput. Syst.},
month = {dec},
articleno = {39},
numpages = {26}
}

@inproceedings{he2023txphishscope,
  title={TxPhishScope: Towards Detecting and Understanding Transaction-based Phishing on Ethereum},
  author={He, Bowen and Chen, Yuan and Chen, Zhuo and Hu, Xiaohui and Hu, Yufeng and Wu, Lei and Chang, Rui and Wang, Haoyu and Zhou, Yajin},
  booktitle={Proceedings of the 2023 ACM SIGSAC Conference on Computer and Communications Security},
  pages={120--134},
  year={2023}
}

@article{li2023towards,
  title={Towards understanding and characterizing the arbitrage bot scam in the wild},
  author={Li, Kai and Guan, Shixuan and Lee, Darren},
  journal={Proceedings of the ACM on Measurement and Analysis of Computing Systems},
  volume={7},
  number={3},
  pages={1--29},
  year={2023},
  publisher={ACM New York, NY, USA}
}

@misc{defihacklabs,
  title={Transaction Simulation Spoofing: A New Threat in Web3},
  howpublished={https://defihacklabs.substack.com/p/transaction-simulation-spoofing-a}
}

@inproceedings{chen2025dissecting,
    title={Dissecting Payload-based Transaction Phishing on Ethereum},
    author={Chen, Zhuo and Hu, Yufeng and He, Bowen and Luo, Dong and Wu, Lei and Zhou, Yajin},
    booktitle={Network and Distributed Systems Security (NDSS) Symposium},
    year={2025}
}

@inproceedings{li2023understanding,
  title={Understanding the Cryptocurrency Free Giveaway Scam Disseminated on Twitter Lists},
  author={Li, Kai and Lee, Darren and Guan, Shixuan},
  booktitle={2023 IEEE International Conference on Blockchain (Blockchain)},
  pages={9--16},
  year={2023},
  organization={IEEE}
}

@inproceedings{torres2019art,
  title={The art of the scam: Demystifying honeypots in ethereum smart contracts},
  author={Torres, Christof Ferreira and Steichen, Mathis and others},
  booktitle={28th USENIX Security Symposium (USENIX Security 19)},
  pages={1591--1607},
  year={2019}
}

@inproceedings{chen2020honeypot,
  title={Honeypot contract risk warning on ethereum smart contracts},
  author={Chen, Weili and Guo, Xiongfeng and Chen, Zhiguang and Zheng, Zibin and Lu, Yutong and Li, Yin},
  booktitle={2020 IEEE International Conference on Joint Cloud Computing},
  pages={1--8},
  year={2020},
  organization={IEEE}
}

@misc{me:polygon,
title = {Polygon},
howpublished = {\url{https://www.polygon.com/

}}
}

@misc{me:avax,
title = {AVAX Network},
howpublished = {\url{https://www.avax.network/
}}
}

@misc{me:bsc,
title = {BNB Chain},
howpublished = {\url{https://www.bnbchain.org/en/bnb-smart-chain}}
}

@inproceedings{guan24ccs,
  title={Characterizing Ethereum address poisoning attack},
  author={Guan, Shixuan and Li, Kai},
  booktitle={Proceedings of the 2024 on ACM SIGSAC Conference on Computer and Communications Security},
  pages={986--1000},
  year={2024}
}

@misc{me:trust,
title = {Trust wallet},
howpublished = {\url{https://trustwallet.com/}}
}

@misc{me:coinbase,
title = {Coinbase},
howpublished = {\url{https://www.coinbase.com/}}
}

@inproceedings{perez2021smart,
  title={Smart contract vulnerabilities: Vulnerable does not imply exploited},
  author={Perez, Daniel and Livshits, Benjamin},
  booktitle={30th USENIX Security Symposium (USENIX Security 21)},
  pages={1325--1341},
  year={2021}
}

@misc{tsuchiya2025blockchainaddresspoisoning,
      title={Blockchain Address Poisoning}, 
      author={Taro Tsuchiya and Jin-Dong Dong and Kyle Soska and Nicolas Christin},
      year={2025},
      eprint={2501.16681},
      archivePrefix={arXiv},
      primaryClass={cs.CR},
      url={https://arxiv.org/abs/2501.16681}, 
}

@INPROCEEDINGS{xia20covidscams,
  author={Xia, Pengcheng and Wang, Haoyu and Luo, Xiapu and Wu, Lei and Zhou, Yajin and Bai, Guangdong and Xu, Guoai and Huang, Gang and Liu, Xuanzhe},
  booktitle={2020 APWG Symposium on Electronic Crime Research (eCrime)}, 
  title={Don’t Fish in Troubled Waters! Characterizing Coronavirus-themed Cryptocurrency Scams}, 
  year={2020},
  volume={},
  number={},
  pages={1-14},
  doi={10.1109/eCrime51433.2020.9493255}
}

\appendix
\section{Ethical Consideration}
For those phishing contracts and contract deployer addresses remain unlabelled on the four blockchains explorers, we have contacted the corresponding explorers to report the phishing addresses to protect cryptocurrency users. Meanwhile, we have also reported the transaction simulation phishing threat and the UI design flaws of presenting misleading and inaccurate simulation results to the crypto wallet developers. We also recommend the proposed countermeasures to them for mitigation consideration. In addition, we are currently deploying our detection system to continuously identify and report phishing addresses involved in this phishing activity. Finally, in this paper, we also tried our best to protect the anonymity of the victims and attackers by shortening their addresses presented in Sec.~\ref{sec:result}. Though the data we collected from the blockchains are already part of the public blockchain ledger, we also discarded them after accomplishing the paper's analysis and writing.


\section{Suspicious Transfer Detection Algorithm}
\label{sec:algo}
Algorithm~\ref{algo:detection} is developed for detecting suspicious transfers from a contract's TAC. In summary, the algorithm starts by searching all \texttt{JUMPI} instructions in the contract's TAC and then analyzing their successor basic blocks to search suspcious transfer patterns, as well as the dependency of the control variable referenced in the \texttt{JUMPI} instruction. 
\begin{algorithm}
\caption{The phishing contract matching algorithm.}
\label{algo:detection}
\begin{algorithmic}[1] 
    \State \textbf{input:} the contract's TAC
    \State \textbf{branch\_conditions} $\gets$ \{CALLER, GAS, GASPRICE, TIMESTAMP, NUMBER\}
    \State condition\_1 $\gets$ \{(\texttt{"caller receives"}, \texttt{"more than callvalue"})\}
    \State condition\_2 $\gets$ \{(\texttt{"an external address receives"}, \texttt{"callvalue"})\}
    \ForAll{(function\_id, tac\_function)}
        \State tac\_function.build\_use\_def\_graph()
        \ForAll{basic\_block \textbf{in} tac\_function.blocks}
            \ForAll{st \textbf{in} basic\_block.statements}
                \If{st \textbf{is} TAC\_JUMPI \textbf{and} st.condition $\in$ branch\_conditions}
                    \State start\_block $\gets$ first block in tac\_function.blocks where block.id = st.block\_id
                    \If{start\_block $\neq$ NULL \textbf{and} number of successors of start\_block = 2}
                        \State succ $\gets$ successor blocks of start\_block
                        \If{find(succ[0], condition\_1) \textbf{and} find(succ[1], condition\_2)}
                            \State \textbf{print}(st.condition, \texttt{"True"})
                        \ElsIf{find(succ[0], condition\_2) \textbf{and} find(succ[1], condition\_1)}
                            \State \textbf{print}(st.condition, \texttt{"True"})
                        \Else
                            \State \textbf{print}(st.condition, \texttt{"False"})
                        \EndIf
                    \EndIf
                \EndIf
            \EndFor
        \EndFor
    \EndFor
\end{algorithmic}
\end{algorithm}

\section{Additional Wallet Testing Screenshots}
\label{sec:screenshots}
\begin{figure*}[t]
\centering

\begin{minipage}[t]{0.48\textwidth}
\centering
\subfloat[Transaction simulation result displayed on Backpack]{
  \includegraphics[width=\linewidth]{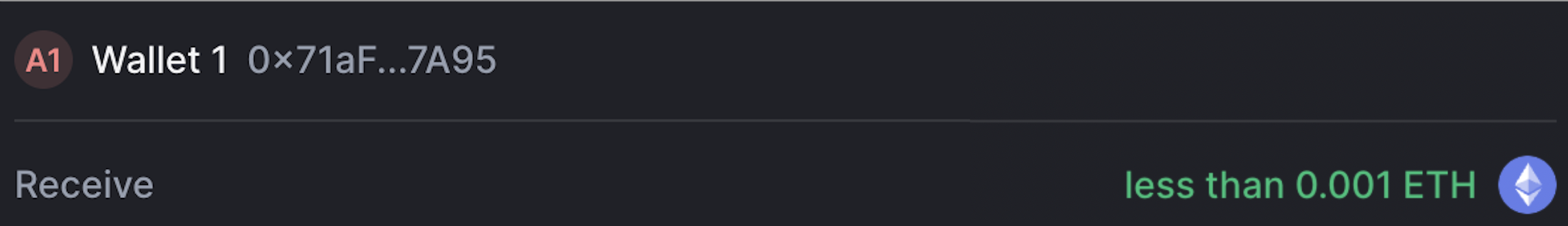}
}\par\vspace{0.6em}

\subfloat[Transaction simulation result displayed on Backpack Core]{
  \includegraphics[width=\linewidth]{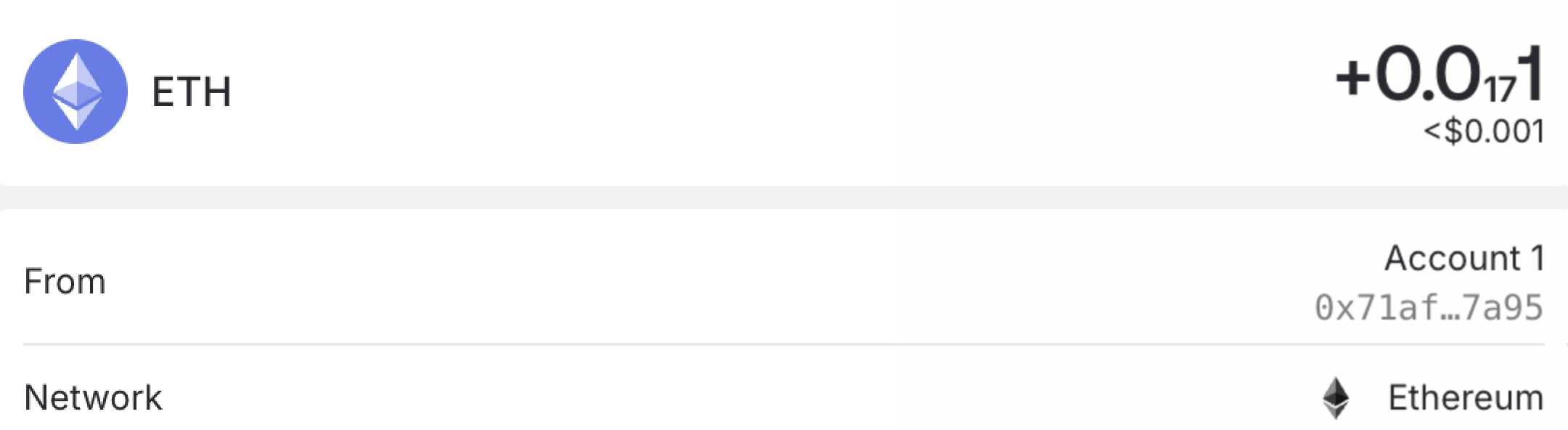}
}\par\vspace{0.6em}

\subfloat[Transaction simulation result displayed on Ctrl]{
  \includegraphics[width=\linewidth]{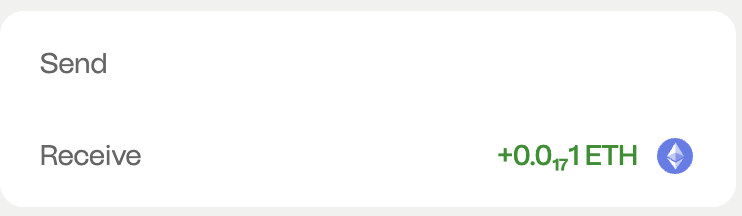}
}\par\vspace{0.6em}

\subfloat[Transaction simulation result displayed on OneKey]{
  \includegraphics[width=\linewidth]{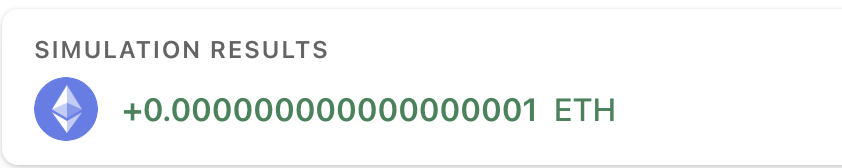}
}
\end{minipage}
\hfill
\begin{minipage}[t]{0.48\textwidth}
\centering
\subfloat[Transaction simulation result displayed on Phantom]{
  \includegraphics[width=\linewidth]{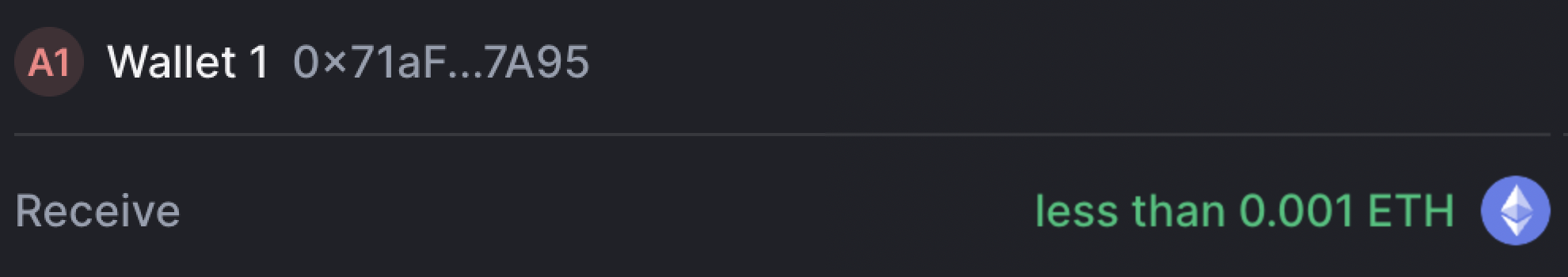}
}\par\vspace{0.6em}

\subfloat[Transaction simulation result displayed on Rabby]{
  \includegraphics[width=\linewidth]{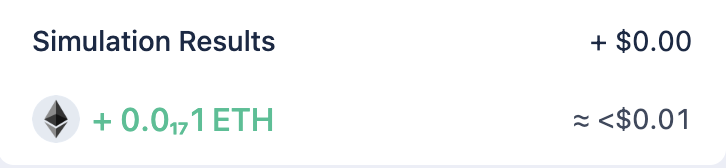}
}\par\vspace{0.6em}

\subfloat[Transaction simulation result displayed on TokenPocket]{
  \includegraphics[width=\linewidth]{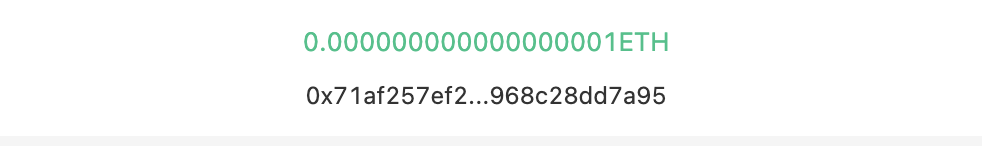}
}\par\vspace{0.6em}

\subfloat[Transaction simulation result displayed on Nest]{
  \includegraphics[width=\linewidth]{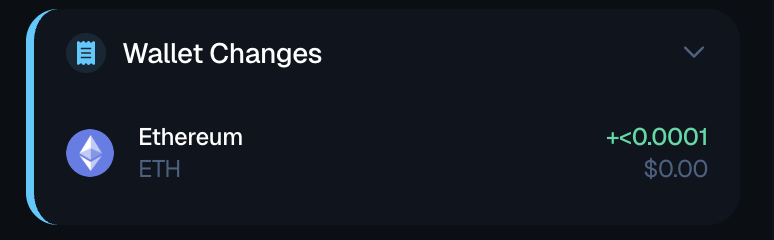}
}
\end{minipage}

\caption{Additional testing screenshots of the transaction simulation results displayed on different wallets' UI when interacting with the phishing contract.}
\label{fig:appendix}
\end{figure*}

Fig.~\ref{fig:appendix} shows the screenshots of transaction simulation results displayed on 8 wallets when interacting with a phishing contract that returns 1 wei.

\section{Justification of Exploitable Dynamic Variables}
\label{sec:justify}
Table~\ref{tab:variables} summarizes the full list of various property variables in the EVM-compatible blockchains. For each variable, we provide a detailed reason on why it can or cannot be abused to implement a phishing contract to exploit the transaction simulation service.

\begin{table*}[!htbp]
\caption{The list of property variables in solidity.}
\label{tab:variables}
\resizebox{1.0\textwidth}{!}{%
\begin{tabular}{ccccc}
\hline
\textbf{Class}               & \textbf{Variable Name} & \textbf{Dynamic?} & \textbf{Can be Abused?} & \textbf{Reason}                                                                             \\ \hline
\multirow{10}{*}{Block}      & blockhash              & yes               & no                      & unpredictable value, offering no deterministic control over the execution path              \\
                             & block.basefee          & yes               & no                      & unpredictable value, offering no deterministic control over the execution path              \\
                             & block.blobbasefee      & yes               & no                      & unpredictable value, offering no deterministic control over the execution path              \\
                             & block.chainid          & no                & no                      & a deterministic number for each chain (mainnet = 1)                                         \\
                             & block.coinbase         & yes               & no                      & unpredictable value, offering no deterministic control over the execution path              \\
                             & block.difficulty       & no                & no                      & fixed at 0 in the mainnet after transitioning to PoS                                        \\
                             & block.gaslimit         & no                & no                      & a deterministic number for each chain (mainnet = 60 million)                                \\
                             & block.number           & yes               & yes                     & monotonically increase, offering a deterministic control over the execution path            \\
                             & block.prevrandao       & yes               & no                      & unpredictable value, attacker has no deterministic control over the execution path          \\
                             & block.timestamp        & yes               & yes                     & monotonically increase, offering a deterministic control over the execution path            \\ \hline
\multirow{8}{*}{Transaction} & gasleft                & yes               & yes                     & adjustable to both transaction simulator or users, can be abused to alter execution path    \\
                             & msg.data               & no                & no                      & cannot be adjusted                                                                          \\
                             & msg.sender             & no                & no                      & cannot be adjusted                                                                          \\
                             & msg.sig                & no                & no                      & cannot be adjusted                                                                          \\
                             & msg.value              & no                & no                      & cannot be adjusted                                                                          \\
                             & blobhash               & no                & no                      & cannot be adjusted                                                                          \\
                             & tx.gasprice            & yes               & yes                     & adjustable to both transaction simulator or users, can be abused to alter execution path    \\
                             & tx.origin              & no                & no                      & cannot be adjusted                                                                          \\ \hline
Smart Contract               & storage                & yes               & yes                     & can be manipulated through sending concurrent transactions to frontrun transaction simlator \\ \hline
\end{tabular}
}
\end{table*}

\end{document}